\documentclass[aps,pra,eqsecnum,showpacs,superscriptaddress]{revtex4}
\usepackage{graphicx,color}
\usepackage{amsmath}
\usepackage{amssymb}
\usepackage{bm}

\renewcommand{\Im}{\mathop{\text{Im}}\nolimits}
\newcommand{\tr}{\mathop{\text{tr}}\nolimits}

\def\uparrow{\delimiter"0222378 }
\def\downarrow{\delimiter"0223379 }

\begin{document}
\title{Entanglement generation by qubit scattering in three dimensions}

\author{Yuichiro Hida}
\affiliation{Department of Physics, Waseda University, Tokyo
169-8555, Japan}

\author{Hiromichi Nakazato}
\affiliation{Department of Physics, Waseda University, Tokyo
169-8555, Japan}

\author{Kazuya Yuasa}
\affiliation{Waseda Institute for Advanced Study, Waseda University,
Tokyo 169-8050, Japan}

\author{Yasser Omar}
\affiliation{CEMAPRE, ISEG, Universidade T\'{e}cnica de Lisboa, P-1200-781 Lisbon, and
SQIG, Instituto de Telecomunica\c{c}\~oes, P-1049-001 Lisbon, Portugal}

\date[]{July 11, 2009}

\begin{abstract}
A qubit (a spin-1/2 particle) prepared in the up state is scattered by local spin-flipping potentials produced by the two target qubits (two fixed spins), both prepared in the down state, to generate an entangled state in the latter when the former is found in the down state after scattering.
The scattering process is analyzed in three dimensions, both to lowest order and in full order in perturbation, with an appropriate renormalization for the latter.
The entanglement is evaluated in terms of the concurrence as a function of the incident and scattering angles, the size of the incident wave packet, and the detector resolution, to clarify the key elements for obtaining an entanglement with high quality.
The characteristics of the results are also discussed in the context of (in)distinguishability of alternative paths for a quantum particle.
\end{abstract}
\pacs{03.67.Bg, 03.65.Nk, 11.10.Gh}

\maketitle

\section{Introduction}
Entanglement plays a crucial role in the field of quantum information and technology~\cite{ref:QIT}, though its acquisition or controlled generation is by no means a trivial matter: it is one of the most peculiar features of quantum theory with no classical analog, and there are several proposals for its generation.
One often makes use of their mutual interaction to make the two quantum systems entangled~\cite{ref:qdots}. 
On the other hand, when they are separated far away and/or when their mutual interaction is considered absent or negligibly weak, one may resort to a third quantum system to make them entangled through its individual interaction with each of them~\cite{ref:cavity,ref:separated,ref:Haroche,ref:photon,ref:scattering,ref:qpfep}. 
For example, one can consider single modes in two cavities with no direct interaction~\cite{ref:cavity}, two spatially separated atoms~\cite{ref:Haroche}, two remote atomic qubits~\cite{ref:photon}, or a system of two magnetic impurities (spins) embedded in a solid~\cite{ref:scattering,ref:qpfep}, and regard an atom, a single cavity mode, a photon, and an electron spin,  respectively, as mediators of interaction between distant quantum parties, in the laboratories.
This kind of scheme with the use of ``entanglement mediators'' has been investigated for simple systems of qubits (quantum two-level systems), usually with the assumption that the strength of  the interaction between the mediator and each quantum system, i.e., the magnitude of the  coupling constants and the interaction durations, are completely under our control~\cite{ref:cavity,ref:separated,ref:Haroche}. 
Conditions under which maximally entangled states are realized are expressed in terms of coupling constants and interaction durations in these cases.

Even though the assumption of complete controllability of such experimental parameters as interaction strength is considered to be legitimate, for example, when the interaction is well controlled by switching on/off the external parameters~\cite{ref:Haroche}, there are still cases in which such an assumption is untenable or its applicability is questionable.
In particular, when the interaction time is not well defined or its definition necessarily requires a resolution in some conceptual issues, like the definition of moments of the beginning and the end of interaction for a particle described by a wave packet with a finite width and scattered by a static potential, we would be forced to treat the process as a quantum mechanical scattering process of a mediator system off the target, where additional (internal, e.g., spin) degrees of freedom are duly taken into account.
In scattering processes, such notions as the first or the last moment of interaction are not considered to be proper issues to be asked, for the interaction is supposed to be gradually turned on and off and the scattering matrix, which describes the transition from the remote past ($t\to-\infty$) to the remote future ($t\to+\infty$), could be the only quantity of physical relevance.
The interaction strength is in a sense automatically and implicitly given and we have no choice of defining or controlling the interaction duration once the initial conditions have been fixed.
It is therefore an interesting and nontrivial matter of physical relevance to examine whether the schemes for entanglement generation or extraction, based on the interaction between the mediator and subsystems, could remain valid and function properly even when one has little controllability on such parameters as time.

In this paper, a three-dimensional scattering process, in which a qubit (a spin-1/2 particle, playing a role of mediator) is scattered off a fixed target composed of two qubits by spin-flipping delta-shaped potentials, is considered to examine the ability of obtaining an entangled state in the target system when a spin flip has been confirmed in the final state of the mediator qubit.
The same setup has already been considered, but essentially only in one (spatial) dimension, to generate a maximally entangled state in the two-qubit system~\cite{ref:scattering,ref:qpfep}. 
It is shown that a maximally entangled state is obtained if one can properly tune the interaction strengths.
On the other hand, if it is treated as a scattering process in one (spatial) dimension, the entanglement can be enhanced by a resonant scattering when the incident momentum of the mediator (or the distance between two target qubits) has been properly chosen in the initial setup~\cite{ref:qpfep}.
Notice that the treatments of the scattering processes so far are not considered to be completely satisfactory because in one dimension, there would be no way to incorporate such important physical parameters as incident and scattering angles, the lateral size of the wave packet and the detector resolution (e.g., an aperture of detector mouth).
The purpose of this paper is to take these elements into account in the scattering process and to clarify the dependence of the resulting entanglement on these parameters.

The paper is organized as follows.
The Hamiltonian for our three-dimensional scattering process is presented in section Sec.~\ref{sec:setup}.  
The scattering matrix (S matrix), which describes the transition of the mediator qubit from a given initial state to the final state, is then introduced in Sec.~\ref{sec:Smatrix}.  
If the scattered qubit is detected with its spin flipped, one knows that the target system of two qubits is in an entangled state, provided that the both target qubits have been prepared in the down state and the interaction preserves the total number of spins-up.
The degree of entanglement is measured in terms of the concurrence and the concurrence is expressed in terms of the S-matrix elements.
In Sec.~\ref{sec:LowestOrder}, the S-matrix elements and the concurrence are calculated to the lowest order and the dependence of the entanglement on physical parameters and its characteristics are discussed.
In this case, the (in)distinguishability of mediator's alternative paths to the detector is shown to be closely connected with the values of concurrence and therefore with the degree of entanglement obtained.
When one tries to evaluate higher-order contributions and goes to higher-order terms in perturbation theory, one realizes that the delta-shaped potentials in three dimensions bring about (ultraviolet) divergences and a proper treatment of such divergences (renormalization) is required~\cite{ref:BegFurlong-Huang,ref:Jackiw}.
We follow the prescription proposed by Jackiw~\cite{ref:Jackiw} to treat delta-shaped potentials and finite S-matrix elements are obtained in Sec.~\ref{sec:higherorder} in full order in perturbation theory by properly introducing counter terms.
The resulting concurrence is examined and its characteristics are discussed in comparison with its lowest-order counterparts.
The final Section~\ref{sec:summary} is devoted to the summary and outlook.
The details of the calculations that are too involved to be presented in the text and some related aspects are shown in Appendices~\ref{app:integralC} and~\ref{app:PQR}.

\section{Setup of the Problem}
\label{sec:setup}
\begin{figure}[b]
\includegraphics[width=0.31\textwidth]{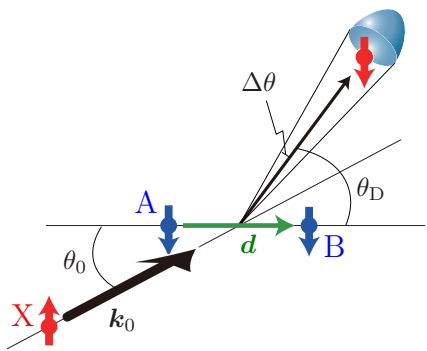}
\caption{(Color online) Qubit X is scattered by qubits A and B and is detected by a detector with a finite resolution.}
\label{fig:Setup}
\end{figure}

The setup is sketched in Fig.~\ref{fig:Setup}.
We try to make two qubits A and B, both prepared in the down state $|\downarrow\downarrow\rangle_\text{AB}$, entangled.
Notice that the initial state of A+B is separable and not entangled at all,  $|\downarrow\downarrow\rangle_\text{AB}\equiv|\downarrow\rangle_\text{A}\otimes|\downarrow\rangle_\text{B}$.
Assume that these qubits are fixed at positions $-{\bm d}/2$ and ${\bm d}/2$, and there is no mutual interaction between them.
In order to make them entangled, another qubit X is prepared in the up state $|\uparrow\rangle_\text{X}$ and is projected on qubits A and B\@.
Qubit X is then scattered by spin-flipping delta-shaped potentials produced by qubits A and B and is finally detected by a spin-sensitive detector.
This physical process can be described by the total Hamiltonian
\begin{equation}
H=\frac{{\bm p}^2}{2m}
+g\delta^3({\bm r}+{\bm d}/2)
\bigl({\bm\sigma}^\text{(X)}\cdot{\bm\sigma}^\text{(A)}\bigr)
+g\delta^3({\bm r}-{\bm d}/2)
\bigl({\bm\sigma}^\text{(X)}\cdot{\bm\sigma}^\text{(B)}\bigr),
\label{eq:H}
\end{equation}
where $\bm p$ and $\bm r$ are the momentum and the position of qubit X of mass $m$ and ${\bm\sigma}^\text{(J)}$ (${\rm J}={\rm X, A, B}$) the Pauli matrices acting on qubit J\@.
The interaction between X and A(B) is described by the spin-exchange operator multiplied by a short-range (delta-shaped) potential with strength $g$ and the energy difference between the up and down states is assumed to be neglected (or the same for all qubits and therefore neglected). 
We shall treat two qubits A and B symmetrically, for simplicity.

Since the Hamiltonian $H$ preserves the total number of ups ($\uparrow$) among the three qubits X, A, and B, if we find qubit X in the spin-flipped (i.e., down) state at the detector after scattering, we are sure that one of the qubits, A \textit{or} B, must be in a spin-flipped state, that is, system A+B is either in $|\uparrow\downarrow\rangle_{\rm AB}$ \textit{or} in $|\downarrow\uparrow\rangle_{\rm AB}$.
If there is no way to judge which spin has been flipped during the scattering by qubit X, the state of qubits A and B is certainly in their superposed state, that is, an entangled state, like 
$\alpha|\uparrow\downarrow\rangle_\text{AB}+\beta|\downarrow\uparrow\rangle_\text{AB}$. 
This argument is valid when the interaction strength can be freely adjusted by some external parameters.
Indeed, it has been shown that a maximally entangled state can be extracted by properly tuning the interaction strength, that is, [the coupling constant]$\times$[interaction duration (time)].

It is, however, to be noticed that when the spatial degrees of freedom of the particles and local potentials are incorporated, as in (\ref{eq:H}), to describe the interactions between qubits X and A(B) as a scattering process, the situation would change drastically:
The state finally be extracted can no longer be a pure state, because it shall be given as a reduced density matrix after being traced over possible spatial or momentum degrees of freedom of X within the detector resolution and therefore it becomes mixed in general.
Furthermore, it is not at all trivial whether we can still expect an entangled state with high quality to be extracted, since we would have no direct control on the interaction time in the scattering process.
(We assume adiabatic switchings of interaction and no (abrupt) changes of the parameters in the Hamiltonian are considered to occur.)
Even if we could deliberately choose the incident momentum of qubit X, so as to adjust effectively the interaction strength, since a quantum particle with a less momentum spread is necessarily accompanied with a longer wave packet in space and the interaction region or range cannot always be given precisely (except for some ideal cases), it is almost impossible or meaningless to talk about the precise moments of particular events in the scattering process.
Stated differently, the quantity of physical relevance in the scattering problems is the scattering (S) matrix elements, which describe the transition from the remote past $t\to-\infty$ to the remote future $t\to+\infty$ under the Hamiltonian $H$.
This is one of the motivations of the previous works where the ability of extraction of entanglement has been examined in the context of the quantum mechanical scattering process in one spatial dimension~\cite{ref:scattering,ref:qpfep}. 
Here the problem is generalized and extended to three spatial dimensions, where a much more rich variety of physical parameters are expected to play roles.

Let the incident qubit X be described by a Gaussian wave packet, with a central momentum $\hbar{\bm k}_0$ and a spatial width $w$.
Qubit X is sent exactly towards the midpoint of target qubits A and B, and the normalized initial state of the total system reads as
\begin{equation}
|\psi_0\rangle
=\int d^3{\bm k}\,\psi_0 ({\bm k})
|{\bm k}\uparrow\downarrow\downarrow\rangle_\text{XAB},\qquad
\psi_0 ({\bm k})
=\left(\frac{2w^2}{\pi}\right)^{3/4}e^{-w^2({\bm k}-{\bm k}_0)^2},
\end{equation}
where $|{\bm k}\uparrow\downarrow\downarrow\rangle_\text{XAB}\equiv|{\bm k}\uparrow\rangle_\text{X}\otimes|\downarrow\downarrow\rangle_\text{AB}$.
A spin-sensitive detector, placed far away from the target detects qubit X scattered in the direction ${\hat{\bm k}}_\text{D}$ seen from the origin (scattering center) with an opening angle $\Delta\theta$.
The detector is assumed to be indifferent to other quantities than spin degrees of freedom of qubit X and we are interested in only those events in which scattered qubit X has been found in the spin-flipped (i.e., down) state $|\downarrow\rangle_\text{X}$.
When qubit X has been found in $|\downarrow\rangle_\text{X}$ in the detector that covers the solid angle
\begin{equation}
\Delta\Omega=\int_0^{2\pi}d\varphi\int_0^{\Delta\theta}d\theta\sin\theta=	4\pi\sin^2(\Delta\theta/2)
\end{equation}
around direction $\hat{\bm k}_\text{D}$, the state of the target system A+B is given by the reduced density matrix of the form
\begin{equation}
\rho
=\frac{1}{P}\int_{\Delta\Omega}d^3{\bm k}\,
{}_\text{X}\langle{\bm k}\downarrow|S|\psi_0\rangle
\langle\psi_0|S^\dagger|{\bm k}\downarrow\rangle_\text{X}
=\frac{1}{P}
\begin{pmatrix}
0&0&0&0\\
0&a_{11}& a_{12}&0\\
0&a_{12}^*& a_{22}&0\\
0&0&0&0 
\end{pmatrix},
\label{eq:rhoAB}
\end{equation}
where $S$ denotes the S matrix and we take 
$|\uparrow\uparrow\rangle_\text{AB}$, $|\uparrow\downarrow\rangle_\text{AB}$,
$|\downarrow\uparrow\rangle_\text{AB}$, and $|\downarrow\downarrow\rangle_\text{AB}$ as the standard basis for the matrix representation of the density operator $\rho$.
Observe that owing to the conserved quantity, i.e., total number of ups, only those elements relevant to $|\uparrow\downarrow\rangle_\text{AB}$ and $|\downarrow\uparrow\rangle_\text{AB}$ can have nonvanishing values in the current setup.
The normalization constant 
\begin{equation}
P=\tr\!\left\{\int_{\Delta\Omega}d^3{\bm k}\,
{}_\text{X}\langle{\bm k}\downarrow|S|\psi_0\rangle\langle\psi_0|S^\dagger|{\bm k}\downarrow\rangle_\text{X}\right\}
=a_{11}+a_{22}
\end{equation}
is nothing but the probability that this particular event occurs, i.e., the yield.
Given the state $\rho$ for two qubits, the degree of entanglement can be measured in terms of its concurrence $C(\rho)$, which reads, for the above $\rho$ in (\ref{eq:rhoAB}), as~\cite{ref:concurrence}
\begin{equation}
C(\rho)=\frac{2|a_{12}|}{a_{11}+a_{22}}.
\end{equation}

\section{Scattering Matrix}
\label{sec:Smatrix}
We proceed to the calculation of the matrix elements $a_{ij}$ in (\ref{eq:rhoAB}).
The scattering matrix $S$ is defined in the following limits
\begin{equation}   
S=\mathop{\lim_{t\to+\infty}}_{t'\to-\infty}e^{iH_0t/\hbar}e^{-iH(t-t')/\hbar}e^{-iH_0t'/\hbar},
\end{equation}
where $H_0={\bm p}^2/2m$ is the free Hamiltonian.
Let $|{\bm k}\zeta\rangle$ be an eigenstate of $H_0$ with $\zeta$ collectively denoting the spin state of three qubits X, A, and B, i.e.,
\begin{equation}
H_0|{\bm k}\zeta\rangle=E_k|{\bm k}\zeta\rangle,
\qquad E_k=\frac{\hbar^2k^2}{2m},
\end{equation}
and $|\Psi_{\bm k}\zeta\rangle$ an eigenstate of the total Hamiltonian $H\equiv H_0+V$ belonging to the same energy $E_k$.
Notice that since the free Hamiltonian $H_0$ is independent of the spin degrees of freedom $\zeta$, the eigenstates are degenerated with respect to $\zeta$.  
The normalized solution of the eigenvalue equation
\begin{equation}
H|\Psi_{\bm k}\zeta\rangle=E_k|\Psi_{\bm k}\zeta\rangle
\end{equation}
is formally given by
\begin{align}
|\Psi_{\bm k}\zeta\rangle
&=|{\bm k}\zeta\rangle+\frac{1}{ E_k-H+i\epsilon}V|{\bm k}\zeta\rangle
\nonumber\\
&=|{\bm k}\zeta\rangle+\frac{1}{ E_k-H_0+i\epsilon}V|\Psi_{\bm k}\zeta\rangle,
\end{align}
\begin{equation}
\langle\Psi_{\bm k}\zeta|\Psi_{{\bm k}'}\zeta'\rangle=\delta^3({\bm k}-{\bm k}')\delta_{\zeta\zeta'}.
\end{equation}
Its coordinate representation reads as
\begin{equation}
\langle{\bm r}|\Psi_{\bm k}\zeta\rangle=\langle{\bm r}|{\bm k}\zeta\rangle-\int d^3{\bm r}'\,G_k({\bm r}-{\bm r}')\frac{2m}{\hbar^2}V({\bm r}')\langle{\bm r}'|\Psi_{\bm k}\zeta\rangle,
\label{eq:rPsi}
\end{equation}
where the retarded Green function $G_k$ is given by
\begin{equation}
G_k({\bm r})
=\frac{\hbar^2}{2m}\int\frac{d^3{\bm q}}{(2\pi)^3}\frac{e^{i{\bm q}\cdot{\bm r}}}{ E_q-E_k-i\epsilon}
=\frac{e^{ikr}}{ 4\pi r}.
\end{equation}
(The above eigenstate satisfies the so-called ``out-going wave" boundary condition for $r\to\infty$.)\ 
These eigenstates $\{|\Psi_{\bm k}\zeta\rangle\}$ form, together with possible bound states $\{|n\chi\rangle\}$ with negative discrete energies $E_{n\chi}<0$, a complete set 
\begin{equation}
\sum_\zeta\int d^3{\bm k}\,|\Psi_{\bm k}\zeta\rangle\langle\Psi_{\bm k}\zeta|+\sum_{n,\chi}|n\chi\rangle\langle n\chi|=1.
\end{equation}

We are now ready to evaluate the S-matrix elements.
Notice first that the following limits are evaluated by making use of the above eigenstates $|\Psi_{\bm k}\zeta\rangle$,
\begin{widetext}
\begin{align}
e^{iHt/\hbar}e^{-iH_0t/\hbar}|{\bm k}\zeta\rangle
&
=\sum_{\zeta'}\int d^3{\bm k}'\,|\Psi_{{\bm k}'}\zeta'\rangle\langle\Psi_{{\bm k}'}\zeta'|{\bm k}\zeta\rangle e^{i(E_{k'}-E_k)t/\hbar}
+\sum_{n,\chi}|n\chi\rangle\langle n\chi|{\bm k}\zeta\rangle e^{i(E_{n\chi}-E_k)t/\hbar}
\nonumber\\
&
\to\begin{cases}
\medskip\displaystyle
|\Psi_{\bm k}\zeta\rangle
&\text{as}\quad t\to-\infty,
\\
\displaystyle
|\Psi_{\bm k}\zeta\rangle
+2\pi i\sum_{\zeta'}\int d^3{\bm k}'\delta(E_k-E_{k'})
|\Psi_{{\bm k}'}\zeta'\rangle\langle\Psi_{{\bm k}'}\zeta'|
V
|{\bm k}\zeta\rangle
&\text{as}\quad t\to+\infty.
\end{cases}
\end{align}
\end{widetext}
These results correctly reflect the fact that the eigenstate $|\Psi_{\bm k}\zeta\rangle$ satisfies the ``out-going wave'' boundary condition and no bound state can survive in the asymptotic regions where the energy conservation is recovered.
The S-matrix element is thus given by
\begin{equation}
\langle{\bm k}'\zeta'|S|{\bm k}\zeta\rangle
=\delta^3({\bm k}'-{\bm k})\delta_{\zeta'\zeta}
-2\pi i\delta(E_{k'}-E_k)\langle{\bm k}'\zeta'|V|\Psi_{\bm k}\zeta\rangle.
\label{eq:Smatrix}
\end{equation}

\section{Born Approximation}
\label{sec:LowestOrder}
To lowest order in the coupling constant $g$, the eigenstate $|\Psi_{\bm k}\zeta\rangle$ in the S-matrix element (\ref{eq:Smatrix}) can be replaced with its free counterpart $|{\bm k}\zeta\rangle$ since $V$ is proportional to $g$.
Up to the first order in $g$, the relevant S-matrix elements read as (subscripts $_\text{X,A,B}$ for the spin states shall be suppressed for notational simplicity in what follows, provided no confusion would arise)
\begin{equation}
\begin{cases}
\medskip
\displaystyle
\langle{\bm k}\downarrow\uparrow\downarrow|S|\psi_0\rangle
=gA({\bm k})+O(g^2),\\
\displaystyle
\langle{\bm k}\downarrow\downarrow\uparrow|S|\psi_0\rangle
=-gA^*({\bm k})+O(g^2),
\end{cases}
\end{equation}
where 
\begin{equation}
A({\bm k})
\equiv-\frac{i}{2\pi^2}\int d^3{\bm k}'\,\psi_0 ({\bm k}')\delta(E_k-E_{k'})e^{i({\bm k}-{\bm k}')\cdot{\bm d}/2}.
\label{eq:A(k)}
\end{equation}
The matrix elements $a_{ij}$ are simply expressed as
\begin{equation}
a_{11}
=a_{22}=g^2\int_{\Delta\Omega}d^3{\bm k}\,|A({\bm k})|^2\equiv a,
\qquad
a_{12}
=-g^2\int_{\Delta\Omega}d^3{\bm k}\,A^2({\bm k})\equiv\tilde a,
\end{equation}
and the concurrence $C$ and the yield $P$ are given by
\begin{equation}
C(\rho)=\frac{|\tilde a|}{ a},\qquad P=2a.
\end{equation}

When the incident wave packet $\psi_0({\bm k})$ is well monochromatized, $wk_0\gg1$, the above quantity $A({\bm k})$ is approximately evaluated analytically.
Since it is expected that the main contributions in the amplitude $A({\bm k})$ in (\ref{eq:A(k)}),
\begin{equation}
A({\bm k})=
-\frac{i}{2\pi^2}
\left(\frac{2w^2}{\pi}\right)^{3/4}
\int_{-\infty}^\infty\frac{ds}{2\pi}
\int d^3{\bm q}\,e^{-w^2{\bm q}^2}
e^{i({\bm k}-{\bm k}_0-{\bm q})\cdot{\bm d}/2}
e^{is(E_k-E_{k_0}-\hbar^2{\bm k}_0\cdot{\bm q}/m-E_q)},
\end{equation}
are due to the integrand with small $|{\bm q}|\lesssim1/w$, we would be allowed to drop the term $E_q$ in the third exponent relative to $E_{k_0}$, for $E_q\lesssim\hbar^2/2mw^2\ll\hbar^2k_0^2/2m=E_{k_0}$.
Then, the integrations over ${\bm q}$ and $s$ are easily performed and we arrive at
\begin{equation}
A({\bm k})\sim-\frac{i}{2\pi^2}
\left(\frac{2\pi}{ w^2}\right)^{3/4}
\sqrt{\frac{mw^2}{2\pi\hbar^2E_{k_0}}}
e^{-w^2k_0^2(E_k/E_{k_0}-1)^2/4}
e^{-i{\bm k}_0\cdot{\bm d}(E_k/E_{k_0}-1)/4}
e^{i({\bm k}-{\bm k}_0)\cdot{\bm d}/2}
e^{-[{\bm d}^2-(\hat{\bm k}_0\cdot{\bm d})^2]/16w^2}
.
\label{eq:Ak}
\end{equation} 
The matrix element $a=a_{11}=a_{22}$ is now reduced to
\begin{equation}
a
\sim\frac{g^2}{4\pi^4}\left(\frac{2\pi}{ w^2}\right)^{3/2}
\frac{mw^2}{2\pi\hbar^2E_{k_0}}
e^{-[{\bm d}^2-(\hat{\bm k}_0\cdot{\bm d})^2]/8w^2}
\int_{\Delta\Omega}d^3{\bm k}\,e^{-w^2k_0^2(E_k/E_{k_0}-1)^2/2}.
\end{equation}
The integration over ${\bm k}$ can be estimated as
\begin{align}
&\int_{\Delta\Omega}d^3{\bm k}\,e^{-w^2k_0^2(E_k/E_{k_0}-1)^2/2}\nonumber\\
&
=\Delta\Omega\,\frac{\sqrt{2m^3E_{k_0}^3}}{\hbar^3}
\int_{-1}^\infty dx\,e^{-w^2k_0^2x^2/2+(1/2)\ln(1+x)}
\sim\Delta\Omega\,\frac{mk_0}{\hbar^2}E_{k_0}
\int_{-\infty}^\infty dx\,e^{-w^2k_0^2x^2/2+x/2}\nonumber\\
&
=\Delta\Omega\,\frac{\sqrt{2\pi m^2}}{\hbar^2w}E_{k_0}\,e^{1/8w^2k_0^2}
\sim\Delta\Omega\,\frac{\sqrt{2\pi m^2}}{w\hbar^2}E_{k_0},
\end{align}
in the lowest order in $1/w^2k_0^2$.
Thus, we obtain
\begin{equation}
a\sim\frac{m^2g^2}{2\pi^3\hbar^4w^2}\,\Delta\Omega\,e^{-[{\bm d}^2-({\hat{\bm k}}_0\cdot{\bm d})^2]/8w^2}.
\end{equation}
We apply the same approximation to the other matrix element $\tilde a=a_{12}$ to get first
\begin{widetext}
\begin{align}
\tilde a
&\sim\frac{g^2}{4\pi^4}
\left(
\frac{2\pi}{w^2}
\right)^{3/2}
\frac{mw^2}{2\pi\hbar^2E_{k_0}}
e^{-[{\bm d}^2-(\hat{\bm k}_0\cdot{\bm d})^2]/8w^2}
\int_{\Delta\Omega}d^3{\bm k}\,
e^{-w^2k_0^2(E_k/E_{k_0}-1)^2/2}
e^{-i{\bm k}_0\cdot{\bm d}(E_k/E_{k_0}-1)/2}
e^{i({\bm k}-{\bm k}_0)\cdot{\bm d}}.
\end{align}
The last integration is similarly approximated as
\begin{align}
&\int_{\Delta\Omega}d^3{\bm k}\,e^{-w^2k_0^2(E_k/E_{k_0}-1)^2/2}
e^{-i{\bm k}_0\cdot{\bm d}(E_k/E_{k_0}-1)/2}
e^{i({\bm k}-{\bm k}_0)\cdot{\bm d}}\nonumber\\
&
=\frac{\sqrt{2m^3E_{k_0}^3}}{\hbar^3}
\int_{\Delta\Omega}d^2\hat{\bm k}
\int_{-1}^\infty dx\,e^{-w^2k_0^2x^2/2
-i({\bm k}_0\cdot{\bm d})x/2+ik_0(\hat{\bm k}\cdot{\bm d})\sqrt{1+x}+(1/2)\ln(1+x)}
e^{-i{\bm k}_0\cdot{\bm d}}\nonumber\\
&
\sim\frac{\sqrt{2\pi m^2}}{\hbar^2w}E_{k_0}
\int_{\Delta\Omega}d^2\hat{\bm k}\,
e^{-[(k_0\hat{\bm k}-{\bm k}_0)\cdot{\bm d}]^2/8w^2k_0^2}
e^{i(k_0\hat{\bm k}-{\bm k}_0)\cdot{\bm d}},
\end{align}
\end{widetext}
which leads to
\begin{equation}
\tilde a\sim
\frac{m^2g^2}{2\pi^3\hbar^4w^2}
e^{-[{\bm d}^2-({\hat{\bm k}}_0\cdot{\bm d})^2]/8w^2}
\int_{\Delta\Omega}d^2\hat{\bm k}\,
e^{-[(k_0\hat{\bm k}-{\bm k}_0)\cdot{\bm d}]^2/8w^2k_0^2}
e^{i(k_0\hat{\bm k}-{\bm k}_0)\cdot{\bm d}}
.
\end{equation}
We finally end up with the following expressions for the concurrence and the yield
\begin{equation}
C(\rho)=\frac{|\tilde a|}{a}
\simeq\frac{1}{\Delta\Omega}
\left|\int_{\Delta\Omega}d^2\hat{\bm k}\, 
e^{ik_0\hat{\bm k}\cdot{\bm d}-[(\hat{\bm k}-{\hat{\bm k}}_0)\cdot{\bm d}]^2/8w^2}\right|,
\label{eq:CBorn}
\end{equation}
\begin{equation}
P=2a
\simeq\frac{m^2g^2}{\pi^3\hbar^4w^2}\,\Delta\Omega\,e^{-[{\bm d}^2-({\hat{\bm k}}_0\cdot{\bm d})^2]/8w^2}.
\label{eq:PBorn}
\end{equation}

The angle integrations over $\hat{\bm k}$ in the concurrence $C$ are numerically performed and the result may be plotted as a surface in three dimensions, swept by a vector, originating from the scattering center and pointing to the detector, i.e., representing the scattering direction of qubit X or the detector direction $\hat{\bm k}_\text{D}$, with its magnitude equal to $C$.
The following figures show sections of such surfaces, which  are cylindrically symmetric with respect to vector $\bm d$, for different values of parameters.

\subsection{Characteristics}
\subsubsection{Cases when the incident wave packet is almost like a plane wave: $w\gg d$}
\label{sssec:lw}
\begin{figure*}
 \begin{center}
\begin{tabular}{l@{\qquad\qquad}l}
(a)&(b)\\[-3.5truemm]
  \includegraphics[height=0.31\textwidth]{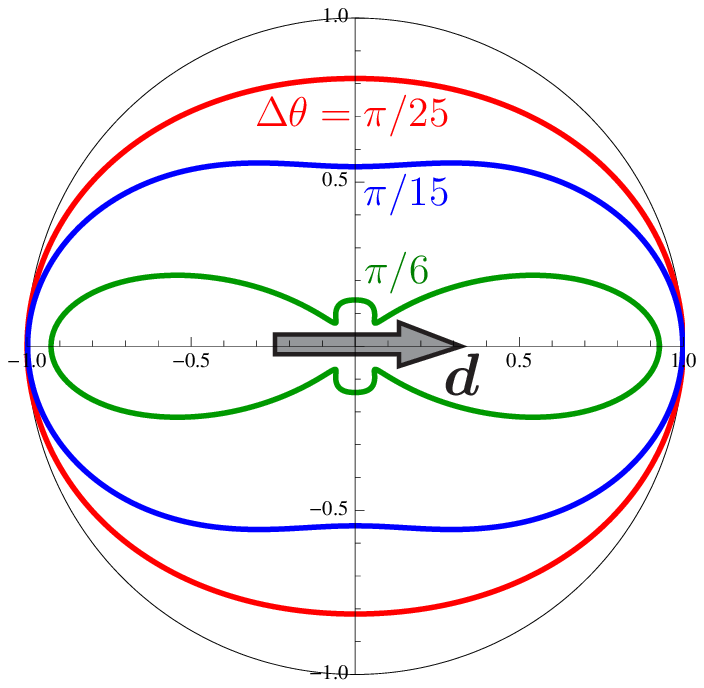}
&
  \includegraphics[height=0.31\textwidth]{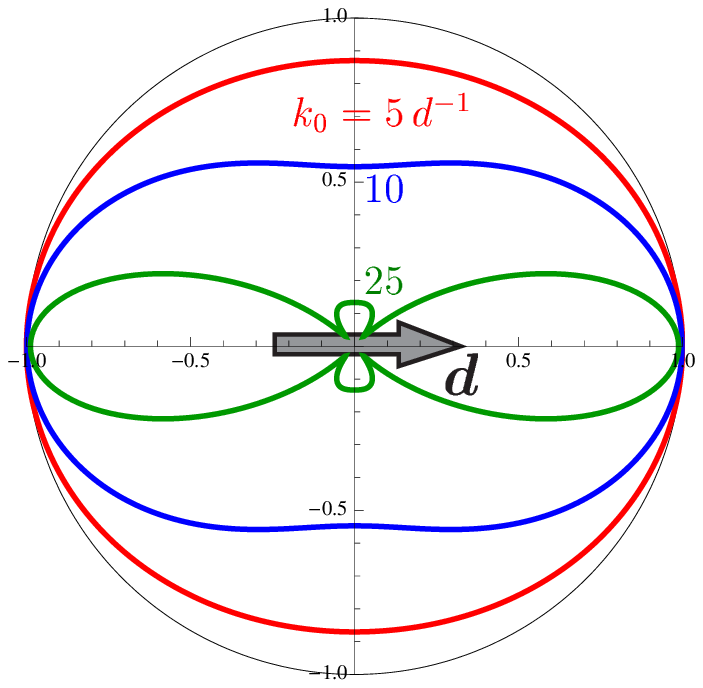}
 \end{tabular}
\end{center}
 \caption{(Color online) Concurrence $C$ in the Born approximation for cases with a large incident wave packet, $w\gg d$, shown in polar coordinates with radius $C$ as a function of the scattering angle $\theta_\text{D}$ relative to the alignment $\bm{d}$ of the target qubits A and B\@.
Parameters are: (a) $\Delta\theta=\pi/25,\,\pi/15,\,\pi/6$ with $k_0d=10$ (dependence on the opening angle of the detector mouth $\Delta\theta$), and 
(b) $k_0d=5,\,10,\,25$ with $\Delta\theta=\pi/15$ (dependence on the incident momentum $k_0$).
The thin-line circles indicate $C=1$ (maximal entanglement).
}
 \label{fig:lwk0hat}
\end{figure*}

Figure~\ref{fig:lwk0hat} shows the concurrence  $C$ generated by the incident wave packet with a large spatial width $w\gg d$.
In this case, (i) the concurrence $C$ is independent of the incident direction $\hat{\bm k}_0\cdot\hat{\bm d}=\cos\theta_0$.
It is clear from these figures that (ii) it depends on in which direction the scattered qubit X is detected (dependence on the scattering direction $\hat{\bm{k}}_\text{D}\cdot\hat{\bm{d}}=\cos\theta_\text{D}$) and it takes the maximal value when qubit X is captured in the direction of $\pm{\bm d}$, i.e., on the line connecting target qubits A and B\@.
The opening angle $\Delta\theta$ of the detector mouth (the detecting resolution) also affects the concurrence $C$: we understand that (iii) the concurrence $C$ is reduced considerably when $\Delta\theta$ is increased, while it still keeps the maximal value in the $\pm{\bm d}$ directions.  
Finally, (iv) the concurrence $C$ is small for a large incident wave momentum $k_0$ or a short wavelength compared with the distance $d$ between the two qubits A and B in the target.

\subsubsection{Cases when the incident wave packet is small: $w\lesssim d$}
\label{sssec:sw}
\begin{figure}
 \begin{center}
  \includegraphics[height=0.31\textwidth]{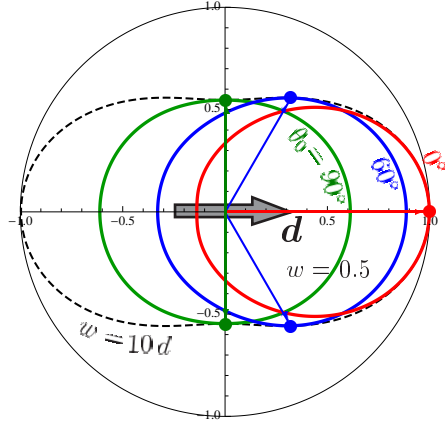}
 \end{center} 
\caption{(Color online) Concurrence $C$ when the incident wave packet is small, $w/d=0.5$, is shown for different incident angles $\theta_0=0^\circ,\,60^\circ,\,90^\circ$ relative to the alignment $\bm{d}$ of qubits A and B, with $k_0d=10$ and $\Delta\theta=\pi/15$.
The concurrence $C$ for a large incident wave packet $w/d=10$ (with the other parameters the same) is also shown (dashed curve) as a reference.
In the directions indicated by the dots (where $\theta_\text{D}=\theta_0$, i.e., in the same direction as the incident direction $\hat{\bm{k}}_0$ and in its cylindrically symmetric directions with respect to $\bm{d}$), the concurrence $C$ takes the same values in both cases $w\gg d$ and $w\lesssim d$.}
\label{fig:swk0hat}
\end{figure}

If the incident wave packet is small compared with the distance between two qubits A and B in the target, i.e., $w\lesssim d$, (v) the concurrence $C$ becomes smaller in general and (vi) the dependence on the incident angle $\theta_0$, which is almost absent for $w\gg d$ in Fig.~\ref{fig:lwk0hat}, appears as seen in Fig.~\ref{fig:swk0hat}.
We observe however that (vii) there are directions where the concurrence $C$ takes the same value as that for the case with a large $w$ shown in Fig.~\ref{fig:lwk0hat}, that is, when qubit X is detected in the same direction as the incident direction $\hat{\bm k}_0$ or in its cylindrically symmetric directions with respect to ${\bm d}$.

\subsubsection{Yield $P$}
\label{sec:CharP}
According to the mathematical expression (\ref{eq:PBorn}), the yield $P$ is independent of the scattering angle $\theta_\text{D}$: no Young-type interference pattern is observed under the condition of the detection of X with its spin flipped.
On the other hand, it depends on the incident direction $\hat{\bm k}_0\cdot{\bm d}$ and becomes maximal when qubit X is injected along the line connecting A and B, i.e., ${\bm k}_0\parallel{\bm d}$.
This may be due to the effective density of the incident probability current felt by the target qubits and the dependence on the incident angle $\theta_0$ disappears when the incident wave packet becomes broad enough $w\gg d$ so that it is seen from the target uniformly spread in space.

\subsection{Conditions for Obtaining Higher Entanglement}
The above characteristics (i)--(vii) of the concurrence $C$ can be understood on the basis of its mathematical expression (\ref{eq:CBorn}).
For the case of large $w\gg d$, it could be further reduced to
\begin{equation}
C(\rho)\simeq\frac{1}{\Delta\Omega}
\left|\int_{\Delta\Omega}d^2\hat{\bm k}\, 
e^{ik_0\hat{\bm k}\cdot{\bm d}}\right|.
\label{eq:CBornlw}
\end{equation}
In order for this quantity to be nonvanishing, the phase of the integrand $k_0\hat{\bm k}\cdot{\bm d}$ should not change considerably within the integration domain $\Delta\Omega$.
That is, the condition under which higher concurrence and therefore higher entanglement is attainable reads as
\begin{equation}
k_0\hat{\bm k}\cdot{\bm d}\,\Bigr|_{\theta_\text{D}+\Delta\theta}^{\theta_\text{D}-\Delta\theta}=2k_0d\sin\theta_\text{D}\sin\Delta\theta\lesssim2\pi
\quad\text{for}\quad w\gg d.
\label{eq:highClw}
\end{equation}
We understand that this condition well describes the characteristics (i)--(iv) mentioned above in Sec.~\ref{sssec:lw}.
Interestingly, it is possible to perform the angle integrations in ({\ref{eq:CBornlw}) analytically (Appendix~\ref{app:integralC}), from which one can derive an approximate expression for the concurrence  
\begin{equation}
C(\rho)\sim1-\frac{1}{2}(k_0d)^2\sin^2\theta_\text{D}\sin^2(\Delta\theta/2)
\label{eqn:Cana}
\end{equation}
when the opening of the detector is sufficiently small $\Delta\theta\ll1$.
It would be evident that the above characteristics are again well explained by this approximate expression.
On the other hand, if the incident wave packet is small $w\lesssim d$, we have to keep the second term $-[(\hat{\bm k}-\hat{\bm k}_0)\cdot{\bm d}]^2/8w^2$ in the exponent in (\ref{eq:CBorn}).
Since this term would entail an exponential reduction of $C$, the condition for keeping a higher concurrence becomes
\begin{equation}
|(\hat{\bm k}-\hat{\bm k}_0)\cdot{\bm d}|\ll w
\quad\text{for}\quad w\lesssim d,
\label{eq:highCsw}
\end{equation}
which explains well the characteristics  (v)--(vii) observed in Sec.~\ref{sssec:sw}.

\subsection{Indistinguishability}
It would be interesting to interpret the above conditions (\ref{eq:highClw}) and (\ref{eq:highCsw}) for higher concurrence $C$ in the context of the (in)distinguishability of the paths taken by particle X\@.
As a general rule in quantum theory, (in)distinguishability of alternatives results in (non)vanishing of quantum interference~\cite{ref:FeynmanLecture}.
Since the concurrence $C$ is proportional to the absolute value of the off-diagonal matrix element $\tilde a=a_{12}$, its value is rather dependent on the information about which qubit A or B has changed its spin state in the scattering process.
If one could obtain such information in principle, there remains no quantum interference between the two alternatives and the off-diagonal elements of the density matrix become vanishingly small.
On the contrary, if there is no way to obtain such information, one can expect a  high quantum interference retained and hence a high entanglement in A and B\@.

Notice also that in the lowest-order perturbation qubit X with its spin flipped has been scattered only and surely once, either by qubit A or B, and the interaction has certainly changed their spin states: there is a direct link between  the information about which qubit has scattered X and that about which spin has been flipped.
Thus we expect that if one could distinguish the two alternative paths of qubit X, originating from qubit A or B, one is able to know which spin has been flipped.
This knowledge would result in a reduction of the off-diagonal matrix elements and therefore of the concurrence $C$.

We understand that the conditions for higher concurrence (entanglement) (\ref{eq:highClw}) and (\ref{eq:highCsw}) can be interpreted as those for the indistinguishability of the two alternative paths from qubit A or B\@.
Indeed, if the incident wave packet is long $w\gg d$ and therefore is approximately considered as a plane wave of wavelength $\lambda_0=2\pi/k_0$, the condition for higher concurrence (\ref{eq:highClw}) is understandable in terms of the resolving power of an optical device. 
It is known in classical optics that the optical device that has an aperture $\Delta\theta$ seen from an object composed of two optical sources with mutual distance $d$ is unable to distinguish the two sources if the condition
\begin{equation}
\Delta\theta\lesssim{\lambda_0\over d\sin\theta_{\rm D}}
\end{equation}         
is satisfied (for $\Delta\theta\ll1$)~\cite{ref:BornWolf}.  
See Fig.~\ref{fig:fraunhofer}.
\begin{figure}
\begin{center}
\includegraphics[width=0.4\textwidth]{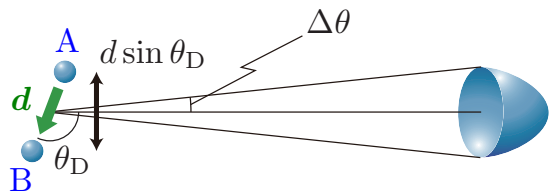}
\end{center}
\caption{(Color online) Resolving power of an optical device and the condition (\ref{eq:highClw}).
The optical device with an aperture $\Delta\theta$ can resolve the separation $d$ if the wavelength $\lambda_0=2\pi/k_0$ is short enough, i.e., when $\lambda_0\lesssim d\,\Delta\theta \sin\theta_{\rm D}$.
Otherwise, it cannot distinguish two sources A and B, the condition of which is nothing but (\ref{eq:highClw}).}
\label{fig:fraunhofer}
\end{figure}
This is essentially the same condition as (\ref{eq:highClw}).
On the other hand, in the opposite case with $w\lesssim d$, since the quantity $|(\hat{\bm k}-\hat{\bm k}_0)\cdot{\bm d}|$ is nothing but the difference in length between the two paths via qubit A or B (see Fig.~\ref{fig:path_length}), if the condition (\ref{eq:highCsw}) is not satisfied, one can determine the path particle X has passed through on its way to the detector, for the difference in the path length is certainly larger than the size of the particle, $w$.
\begin{figure}
 \begin{center}
  \includegraphics[width=0.35\textwidth]{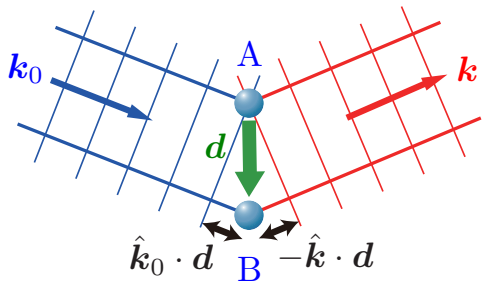}
 \end{center}
 \caption{(Color online) Two paths via A and B have difference in their lengths by $|(\hat{\bm k}_0-\hat{\bm k} )\cdot{\bm d}|$.}
 \label{fig:path_length}
\end{figure}
In this way, the conditions for higher entanglement, in both cases, are interpreted as those for the indistinguishability of the paths taken by X\@.

The one-to-one correspondence between the path of X and the flipped spin A or B also explains why no interference is observed in the yield $P$ (Sec.~\ref{sec:CharP}).
One can know the path of X, either via A or B, by looking at the spin state of A and B after scattering.
This accessibility to the information on the path of X erases the interference between the two alternative paths.

Notice, however, that these relations of the concurrence $C$ and the yield $P$ with the indistinguishability are limited to the lowest-order results and further study is necessary for higher-order terms in perturbation theory, where multiple scatterings, which invalidate the one-to-one correspondence between the knowledge of the particle paths and that of spin flips, are present and a resonant scattering is expected to play a role.

\section{Estimation of Higher-Order Terms}
\label{sec:higherorder}
\subsection{Spin-flipping delta-shaped potential and renormalization}
\label{ssec:renormalization}
The multiple scatterings are absent in the lowest-order terms in perturbation and their effects, including those of resonant scatterings, can be seen only when one goes to the higher-order calculations.
In this case, one realizes that a proper treatment of them is required, for otherwise the result would become trivial, that is, no scattering would occur by the delta-shaped potentials in dimensions greater than one~\cite{ref:BegFurlong-Huang}.
This is a famous issue in quantum theory and Jackiw has proposed a prescription how to deal with such systems~\cite{ref:Jackiw}. According to Jackiw's prescription, we have to renormalize the strength of the delta-shaped potential so that the source term becomes nonvanishing.
In other words, the coupling constant in the Hamiltonian, $g$ in our case, has to be regarded as a bare one and absorb possible divergences arising from the Green function at the origin $G_k({\bm0})$.

In our case, however, another element that was absent at that time, i.e., the spin degrees of freedom, come into play, which would require another care in dealing with higher-order terms.
In this respect, it is important to notice that the interaction of the form $g{\bm\sigma}^\text{(X)}\cdot{\bm\sigma}^\text{(A)}$ inevitably causes another type of interaction in its higher-order terms.
For example, in its second order, a term proportional to the unit operator in spin space, that is not proportional to the original form of the interaction, appears:
\begin{equation}
(
g{\bm\sigma}^\text{(X)}\cdot{\bm\sigma}^\text{(A)}
)^2
=3g^2-2g^2{\bm\sigma}^\text{(X)}\cdot{\bm\sigma}^\text{(A)}.
\end{equation} 
This is easily understood from the fact that the interaction can be written as
\begin{equation}
{\bm\sigma}^\text{(X)}\cdot{\bm\sigma}^\text{(A)}={\cal P}_\text{3XA}-3{\cal P}_\text{1XA},
\end{equation}
where ${\cal P}_\text{3XA}$ and ${\cal P}_\text{1XA}$ are the projection operators on the spin-triplet and singlet spaces, respectively,
\begin{equation}
{\cal P}_\text{3XA}
=\frac{1}{4}(3+{\bm\sigma}^\text{(X)}\cdot{\bm\sigma}^\text{(A)}),
\qquad
{\cal P}_\text{1XA}
=\frac{1}{4}(1-{\bm\sigma}^\text{(X)}\cdot{\bm\sigma}^\text{(A)}),
\end{equation}
\begin{equation}
{\cal P}_\text{3XA}^2={\cal P}_\text{3XA},
\quad
{\cal P}_\text{1XA}^2={\cal P}_\text{1XA},
\quad
{\cal P}_\text{3XA}{\cal P}_\text{1XA}={\cal P}_\text{1XA}{\cal P}_\text{3XA}=0,
\quad
{\cal P}_\text{3XA}+{\cal P}_\text{1XA}=1.
\end{equation}
This means that in any higher-order terms in perturbation, there are only two types of interactions proportional to ${\cal P}_\text{3XA}$ or ${\cal P}_\text{1XA}$ present and we need to renormalize these two terms simultaneously.
That is, we need two counter terms to obtain sensible results.
Let us therefore introduce another bare coupling constant $g'$ and start from a bare interaction Hamiltonian 
\begin{equation}
V({\bm r})
=\delta^3({\bm r}+{\bm d}/2)(g{\bm\sigma}^\text{(X)}\cdot{\bm\sigma}^\text{(A)}+g')
+\delta^3({\bm r}-{\bm d}/2)(g{\bm\sigma}^\text{(X)}\cdot{\bm\sigma}^\text{(B)}+g'),
\end{equation}
which is also written as
\begin{align}
V({\bm r})
&=\delta^3({\bm r}+{\bm d}/2)
[
(g+g'){\cal P}_\text{3XA}+(-3g+g'){\cal P}_\text{1XA}
]
+\delta^3({\bm r}-{\bm d}/2)
[
(g+g'){\cal P}_\text{3XB}+(-3g+g'){\cal P}_\text{1XB}
]\nonumber\\
&\equiv
\frac{\hbar^2}{2m}\delta^3({\bm r}+{\bm d}/2){\cal Q}_\text{XA}
+\frac{\hbar^2}{2m}\delta^3({\bm r}-{\bm d}/2){\cal Q}_\text{XB}.
\end{align}

We can formally solve the equation for the eigenstate given in (\ref{eq:rPsi}) in the coordinate representation $\langle{\bm r}|\Psi_{\bm k}\zeta\rangle$,
\begin{equation}
\langle{\bm r}|\Psi_{\bm k}\zeta\rangle
=\langle{\bm r}|{\bm k}\zeta\rangle
-G_k({\bm r}+{\bm d}/2){\cal Q}_\text{XA}\langle-{\bm d}/2|\Psi_{\bm k}\zeta\rangle
-G_k({\bm r}-{\bm d}/2){\cal Q}_\text{XB}\langle{\bm d}/2|\Psi_{\bm k}\zeta\rangle,
\end{equation}
from which the source terms $\propto\langle\pm{\bm d}/2|\Psi_{\bm k}\zeta\rangle$ can be derived by solving 
\begin{equation}
\begin{pmatrix}
1+G_k({\bm0}){\cal Q}_\text{XB}&G_k({\bm d}){\cal Q}_\text{XA}\\
\noalign{\medskip}
G_k(-{\bm d}){\cal Q}_\text{XB}&1+G_k({\bm0}){\cal Q}_\text{XA}
\end{pmatrix}
\begin{pmatrix}
\langle{\bm d}/2|\Psi_{\bm k}\zeta\rangle\\
\noalign{\medskip}
\langle-{\bm d}/2|\Psi_{\bm k}\zeta\rangle
\end{pmatrix}
=
\begin{pmatrix}
\langle{\bm d}/2|{\bm k}\zeta\rangle\\
\noalign{\medskip}
\langle-{\bm d}/2|{\bm k}\zeta\rangle
\end{pmatrix}.
\end{equation}
After an elementary but a little bit tedious calculation, we formally arrive at
\begin{subequations}
\label{eq:QXABPsi}
\begin{align}
{\cal Q}_\text{XA}\langle-{\bm d}/2|\Psi_{\bm k}\zeta\rangle
&={\cal A}{1\over1-G_k(-{\bm d}){\cal B}G_k({\bm d}){\cal A}}
\,\Bigl(\langle-{\bm d}/2|{\bm k}\zeta\rangle-G_k(-{\bm d}){\cal B}\langle{\bm d}/2|{\bm k}\zeta\rangle\Bigr),
\label{eq:QXAPsi}
\displaybreak[0]\\
{\cal Q}_\text{XB}\langle{\bm d}/2|\Psi_{\bm k}\zeta\rangle
&={\cal B}{1\over1-G_k({\bm d}){\cal A}G_k(-{\bm d}){\cal B}}
\,\Bigl(\langle{\bm d}/2|{\bm k}\zeta\rangle-G_k({\bm d}){\cal A}\langle-{\bm d}/2|{\bm k}\zeta\rangle\Bigr),\label{eq:QXBPsi}
\end{align}
\end{subequations}
where we have defined
\begin{equation}
{\cal A}\equiv{\cal Q}_\text{XA}{1\over1+G_k({\bm0}){\cal Q}_\text{XA}},\qquad
{\cal B}\equiv{\cal Q}_\text{XB}{1\over1+G_k({\bm0}){\cal Q}_\text{XB}}.
\label{eq:calAB}
\end{equation}
It would be evident that these expressions (\ref{eq:QXAPsi}) and (\ref{eq:QXBPsi}) clearly represent multiple scattering processes in terms of the effective (self-)couplings at qubit A and B, ${\cal A}$ and $\cal B$, and the amplitudes $G_k({\bm d})$ and $G_k(-{\bm d})$, describing particle's propagations, A$\to$B and B$\to$A, respectively.

At this point, we have to recall that in dimensions greater than or equal to two, the Green function at the origin $G_k({\bm0})$ is divergent.
Actually in three dimensions, it diverges linearly with a cutoff $\Lambda$,
\begin{align}
G_k({\bm0})
&=\int\frac{d^3{\bm q}}{(2\pi)^3}\frac{1}{q^2-k^2-i\epsilon}
\nonumber\\
&=\lim_{\Lambda\to\infty}\frac{1}{(2\pi)^2}\int_0^\Lambda dq\,q\left(\frac{1}{q-k-i\epsilon}+\frac{1}{q+k+i\epsilon}\right)
\nonumber\\
&=\lim_{\Lambda\to\infty}\frac{1}{2\pi^2}\Lambda+\Omega_k,\quad\Omega_k\equiv{ik\over4\pi}.\end{align}
Therefore we need appropriate renormalizations of the coupling constants in order to obtain nontrivial scattering amplitudes, even though we also know that, according to the dimensional regularization, there are no ultraviolet divergences in odd dimensions and it gives exactly the same finite term.
Since the above operators ${\cal A}$ and ${\cal B}$ are explicitly evaluated, e.g.,
\begin{widetext}
\begin{align}
{\cal A}&={2m\over\hbar^2}[(g+g'){\cal P}_\text{3XA}+(-3g+g'){\cal P}_\text{1XA}]
{1\over1+(2m/\hbar^2)G_k({\bm0})[(g+g'){\cal P}_\text{3XA}+(-3g+g'){\cal P}_\text{1XA}]}\nonumber\displaybreak[0]\\
&={2m\over\hbar^2}\left({g+g'\over1+2m(g+g')G_k({\bm0})/\hbar^2}{\cal P}_\text{3XA}+{-3g+g'\over1+2m(-3g+g')G_k({\bm0})/\hbar^2}{\cal P}_\text{1XA}\right),
\end{align}
\end{widetext}
and the divergences only appear through these operators ${\cal A}$ and ${\cal B}$ in (\ref{eq:calAB}), the following renormalizations of the bare coupling constants are sufficient to make everything finite and nontrivial,
\begin{subequations}
\begin{gather}
\frac{1}{g_r+g_r'}=\frac{1}{g+g'}+{2m\over\hbar^2}\lim_{\Lambda\to\infty}\frac{1}{2\pi^2}\Lambda
,\\
\frac{1}{-3g_r+g_r'}=\frac{1}{-3g+g'}+{2m\over\hbar^2}\lim_{\Lambda\to\infty}\frac{1}{2\pi^2}\Lambda,
\end{gather}
\end{subequations}
by which the above ${\cal A(B)}$ reads as 
\begin{equation}
{\cal A(B)}={2m\over\hbar^2}\left(
\frac{g_r+g_r'}{1+2m(g_r+g_r')\Omega_k/\hbar^2}{\cal P}_\text{3XA(B)}
+\frac{-3g_r+g_r'}{1+2m(-3g_r+g_r')\Omega_k/\hbar^2}{\cal P}_\text{1XA(B)}
\right).
\end{equation}

\subsection{Multiple scatterings in terms of projection operators}
\label{ssedc:multiplescattering}
Assume that the renormalized interaction takes the spin-exchange form (\ref{eq:H}) with the renormalized coupling constant $g_r$, in other words, we shall set the other coupling constant vanishing $g_r'=0$.
In this case, we have
\begin{align}
{\cal A(B)}&=\frac{2m}{\hbar^2}
\left(
\frac{g_r}{1+2mg_r\Omega_k/\hbar^2}{\cal P}_\text{3XA(B)}
+\frac{-3g_r}{1-6mg_r\Omega_k/\hbar^2}{\cal P}_\text{1XA(B)}
\right)
\nonumber\\
&={2mg_r\over\hbar^2}{1\over(1+\xi_k/3)(1-\xi_k)}({\bm\sigma}^\text{(X)}\cdot{\bm\sigma}^\text{(A(B))}-\xi_k),
\end{align}
where $\xi_k\equiv6mg_r\Omega_k/\hbar^2$, and therefore
\begin{widetext}
\begin{subequations}
\begin{align}
G_k(-{\bm d}){\cal B}G_k({\bm d}){\cal A}
&=G_k^2(d)\left(\frac{2mg_r}{\hbar^2}\right)^2{1\over(1+\xi_k/3)^2(1-\xi_k)^2}({\bm\sigma}^\text{(X)}\cdot{\bm\sigma}^\text{(B)}-\xi_k)({\bm\sigma}^\text{(X)}\cdot{\bm\sigma}^\text{(A)}-\xi_k)\nonumber\\
&\equiv f_k^2[
i{\bm\sigma}^\text{(X)}\cdot({\bm\sigma}^\text{(B)}\times{\bm\sigma}^\text{(A)})-\xi_k{\bm\sigma}^\text{(X)}\cdot({\bm\sigma}^\text{(A)}+{\bm\sigma}^\text{(B)})+{\bm\sigma}^\text{(A)}\cdot{\bm\sigma}^\text{(B)}+\xi_k^2
]
\end{align}
and similarly
\begin{equation}
G_k({\bm d}){\cal A}G_k(-{\bm d}){\cal B}
=f_k^2[
-i{\bm\sigma}^\text{(X)}\cdot({\bm\sigma}^\text{(B)}\times{\bm\sigma}^\text{(A)})-\xi_k{\bm\sigma}^\text{(X)}\cdot({\bm\sigma}^\text{(A)}+{\bm\sigma}^\text{(B)})+{\bm\sigma}^\text{(A)}\cdot{\bm\sigma}^\text{(B)}+\xi_k^2
]
\end{equation}
\end{subequations}
with 
\begin{equation}
f_k\equiv G_k(d){2mg_r\over\hbar^2}{1\over(1+\xi_k/3)(1-\xi_k)},
\qquad
G_k(\pm{\bm d})={e^{ikd}\over4\pi d}\equiv G_k(d).
\end{equation}

In order to evaluate the higher-order terms, which have formally been summed up like 
(\ref{eq:QXABPsi}), one needs to know the powers of the above quantities $G_k^2(d){\cal BA}$ and $G_k^2(d){\cal AB}$.
At this point, observe that the three spin operators ${\bm\sigma}^\text{(X)}\cdot({\bm\sigma}^\text{(B)}\times{\bm\sigma}^\text{(A)})$, ${\bm\sigma}^\text{(X)}\cdot({\bm\sigma}^\text{(B)}+{\bm\sigma}^\text{(A)})$ and ${\bm\sigma}^\text{(A)}\cdot{\bm\sigma}^\text{(B)}$ form a closed algebra with respect to the symmetrized multiplications among them.
This implies a possibility that there are projection operators made of these three operators, in terms of which the operators $\cal BA$ and $\cal AB$ can be expanded uniquely.
Indeed, one can confirm, after elementary but tedious calculations, that operators defined by
\begin{subequations}
\begin{align}
{\cal P}_s(\gamma)
&\equiv
{s\over2}\sqrt{\left({1\over4}+3\gamma\right)\left({1\over4}-\gamma\right)}{\bm\sigma}^\text{(X)}\cdot({\bm\sigma}^\text{(B)}\times{\bm\sigma}^\text{(A)})
+{1\over2}\left({1\over4}-\gamma\right){\bm\sigma}^\text{(X)}\cdot({\bm\sigma}^\text{(A)}+{\bm\sigma}^\text{(B)})
+\gamma{\bm\sigma}^\text{(A)}\cdot{\bm\sigma}^\text{(B)}+{3\over4},\\
{\cal Q}_s(\gamma)
&\equiv
-{s\over2}\sqrt{\left({1\over4}+3\gamma\right)\left({1\over4}-\gamma\right)}{\bm\sigma}^\text{(X)}\cdot({\bm\sigma}^\text{(B)}\times{\bm\sigma}^\text{(A)})
-{1\over2}\left({1\over4}-\gamma\right){\bm\sigma}^\text{(X)}\cdot({\bm\sigma}^\text{(A)}
+{\bm\sigma}^\text{(B)})-\gamma{\bm\sigma}^\text{(A)}\cdot{\bm\sigma}^\text{(B)}+{1\over4},\\
{\cal R}_\pm
&\equiv\pm{1\over6}[
{\bm\sigma}^\text{(X)}\cdot({\bm\sigma}^\text{(A)}+{\bm\sigma}^\text{(B)})+{\bm\sigma}^\text{(A)}\cdot{\bm\sigma}^\text{(B)}
]+\frac{1}{2},
\end{align}
\end{subequations}
with $s^2=1$ and $\gamma$ an arbitrary parameter, satisfy the following projective relations
\begin{gather}
{\cal P}_s(\gamma)+{\cal Q}_s(\gamma)=1,\quad
{\cal P}_s(\gamma){\cal Q}_s(\gamma)={\cal Q}_s(\gamma){\cal P}_s(\gamma)=0,\quad
{\cal P}_s^2(\gamma)={\cal P}_s(\gamma),
\label{eq:P1}
\\
{\cal R}_++{\cal R}_-=1,\quad
{\cal R}_+{\cal R}_-={\cal R}_-{\cal R}_+=0,\quad
{\cal R}_+^2={\cal R}_+.
\label{eq:P2}
\end{gather}
Notice that the operators ${\cal R}_\pm$ are not necessarily orthogonal to ${\cal P}_s(\gamma)$ and ${\cal Q}_s(\gamma)$.
In fact,
\begin{equation}
{\cal P}_s(\gamma){\cal R}_+={\cal R}_+{\cal P}_s(\gamma)={\cal R}_+,
\label{eq:P3}
\end{equation}
so that ${\cal R}_+$ is orthogonal to ${\cal Q}_s(\gamma)$, but
\begin{equation}
{\cal Q}_s(\gamma){\cal R}_-={\cal R}_-{\cal Q}_s(\gamma)={\cal Q}_s(\gamma),
\end{equation}
which means that ${\cal R}_-$ is not orthogonal to ${\cal P}_s(\gamma)$ and  
\begin{equation}
{\cal P}_s(\gamma){\cal R}_-={\cal R}_-{\cal P}_s(\gamma)={\cal P}_s(\gamma)-{\cal R}_+\equiv{\cal Q}_s^\perp,
\end{equation}
that is orthogonal to ${\cal Q}_s(\gamma)$, as it should be.
The meanings of these projection operators are exposed in Appendix~\ref{app:PQR}\@.

Three operators ${\cal P}_s(\gamma)$, ${\cal Q}_s(\gamma)$, and ${\cal R}_+$ may be used to uniquely expand the relevant operators $\cal BA$ and $\cal AB$.
This is indeed possible and one finds that
\begin{equation}
G_k^2(d){\cal BA}=\alpha{\cal P}_\mp(\bar\gamma)+\beta{\cal R}_++\delta{\cal Q}_\mp(\bar\gamma),\quad
G_k^2(d){\cal AB}=\alpha{\cal P}_\pm(\bar\gamma)+\beta{\cal R}_++\delta{\cal Q}_\pm(\bar\gamma)
\end{equation}
with appropriately chosen parameters
\begin{equation}
\bar\gamma=\frac{1}{12}\left(
1\pm2i{1+\xi_k\over\sqrt{3-(1+\xi_k)^2}}
\right)
\label{eq:gammabar}
\end{equation}
and
\begin{equation}
\alpha=f_k^2\bigl(1\mp i\sqrt{3-(1+\xi_k)^2}\bigr)^2,\quad
\beta=2f_k^2\bigl(1-2\xi_k\pm i\sqrt{3-(1+\xi_k)^2}\bigr),\quad
\delta=f_k^2\bigl(1\pm i\sqrt{3-(1+\xi_k)^2}\bigr)^2.
\end{equation}
The projective properties of the operators ${\cal P}_s(\gamma)$, ${\cal Q}_s(\gamma)$, and ${\cal R}_+$, shown in (\ref{eq:P1})--(\ref{eq:P3}), easily lead us to
\begin{subequations}
\label{eq:1-BA-AB}
\begin{align}
{1\over1-G_k(-{\bm d}){\cal B}G_k({\bm d}){\cal A}}&={1\over1-\alpha}{\cal P}_\mp(\bar\gamma)+\left(
{1\over1-(\alpha+\beta)}-{1\over1-\alpha}
\right){\cal R}_++{1\over1-\delta}{\cal Q}_\mp(\bar\gamma),
\label{eq:1-BA}
\intertext{and}
{1\over1-G_k({\bm d}){\cal A}G_k(-{\bm d}){\cal B}}&={1\over1-\alpha}{\cal P}_\pm(\bar\gamma)+\left(
{1\over1-(\alpha+\beta)}-{1\over1-\alpha}
\right){\cal R}_++{1\over1-\delta}{\cal Q}_\pm(\bar\gamma).
\label{eq:1-AB}
\end{align}
\end{subequations}

\subsection{S-matrix elements and the concurrence $C$ in full order}
\label{ssec:SandCinfull}
The relevant S-matrix element is now expressed as ($\zeta_\text{AB}=\uparrow\downarrow$ or $\downarrow\uparrow$)
\begin{align}
\langle{\bm k}\downarrow\zeta_\text{AB}|S|\psi_0\rangle
&=\int d^3{\bm k}'\,\langle{\bm k}\downarrow\zeta_\text{AB}|S|{\bm k}'\uparrow\downarrow\downarrow\rangle\psi_0({\bm k}')
\nonumber\\
&=\int d^3{\bm k}'\,\psi_0({\bm k}')(-2\pi i)\delta(E_k-E_{k'})\langle{\bm k}\downarrow\zeta_\text{AB}|V|\Psi_{{\bm k}'}\uparrow\downarrow\downarrow\rangle\nonumber\\
&=\int d^3{\bm k}'\psi_0({\bm k}')(-2\pi i)\delta(E_k-E_{k'})\nonumber\\
&\qquad\qquad\times
{\hbar^2\over2m}\,\Bigl(
\langle{\bm k}|{-{\bm d}/2}\rangle\langle\downarrow\zeta_\text{AB}|{\cal Q}_\text{XA}\langle{-{\bm d}/2}|\Psi_{{\bm k}'}\uparrow\downarrow\downarrow\rangle+\langle{\bm k}|{\bm d}/2\rangle\langle\downarrow\zeta_\text{AB}|{\cal Q}_\text{XB}\langle{\bm d}/2|\Psi_{{\bm k}'}\uparrow\downarrow\downarrow\rangle
\Bigr)\nonumber\\
&=-i\int{d^3{\bm k}'\over(2\pi)^2}\,\psi_0({\bm k}')\delta(E_k-E_{k'})\nonumber\\
&\qquad\qquad\qquad
{}\times\left(
\langle\downarrow\zeta_\text{AB}|{\hbar^2\over2m}{\cal A}{1\over1-G_k^2(d){\cal BA}}|\uparrow\downarrow\downarrow\rangle e^{i({\bm k}-{\bm k}')\cdot{\bm d}/2}
\right.\nonumber\\
&\left.
\qquad\qquad\qquad\qquad
{}-\langle\downarrow\zeta_\text{AB}|{\hbar^2\over2m}{\cal AB}G_k(d){1\over1-G_k^2(d){\cal AB}}|\uparrow\downarrow\downarrow\rangle e^{i({\bm k}+{\bm k}')\cdot{\bm d}/2}
\right.
\nonumber\\
&\left.
\qquad\qquad\qquad\qquad
{}+\langle\downarrow\zeta_\text{AB}|{\hbar^2\over2m}{\cal B}{1\over1-G_k^2(d){\cal AB}}|\uparrow\downarrow\downarrow\rangle e^{-i({\bm k}-{\bm k}')\cdot{\bm d}/2}
\right.
\nonumber\displaybreak[0]\\
&\left.
\qquad\qquad\qquad\qquad
{}-\langle\downarrow\zeta_\text{AB}|{\hbar^2\over2m}{\cal BA}G_k(d){1\over1-G_k^2(d){\cal BA}}|\uparrow\downarrow\downarrow\rangle e^{-i({\bm k}+{\bm k}')\cdot{\bm d}/2}
\right).
\end{align}
By plugging the explicit forms of the inverse operators in terms of the projective ones in (\ref{eq:1-BA-AB}) and evaluating the matrix elements of the three spin operators, ${\bm\sigma}^\text{(X)}\cdot({\bm\sigma}^\text{(B)}\times{\bm\sigma}^\text{(A)})$, ${\bm\sigma}^\text{(X)}\cdot({\bm\sigma}^\text{(B)}+{\bm\sigma}^\text{(A)})$, and ${\bm\sigma}^\text{(A)}\cdot{\bm\sigma}^\text{(B)}$, the relevant matrix elements are calculated, under the same approximation as we have taken in deriving (\ref{eq:Ak}), to be 
\begin{subequations}
\label{eq:kdud-dduSpsi0}
\begin{align}
\langle{\bm k}\downarrow\uparrow\downarrow|S|\psi_0\rangle
&\sim{\cal N}_k\Bigl\{
A({\bm k})[1+f_k^2(1-\xi_k^2)]
+A^*(-{\bm k})[-(1+\xi_k)+(1-\xi_k)^2(\xi_k+3)f_k^2]f_k\nonumber\\
&\quad\qquad
-A^*({\bm k})(1-\xi_k)[-(1+\xi_k)+(1-\xi_k)^2(\xi_k+3)f_k^2]f_k^2
-A(-{\bm k})(1-\xi_k)[1+f_k^2(1-\xi_k^2)]f_k
\Bigr\}\nonumber\\
&={\cal N}_k[A({\bm k})a_k+A^*({\bm k})e^{i{\bm k}\cdot{\bm d}}b_k]
[1-e^{-i{\bm k}\cdot{\bm d}}(1-\xi_k)f_k]
\label{eq:kdudSpsi0}
\intertext{and}
\langle{\bm k}\downarrow\downarrow\uparrow|S|\psi_0\rangle
&\sim{\cal N}_k\Bigl\{
A({\bm k})(1-\xi_k)[-(1+\xi_k)+(1-\xi_k)^2(\xi_k+3)f_k^2]f_k^2
+A^*(-{\bm k})(1-\xi_k)[1+f_k^2(1-\xi_k^2)]f_k
\nonumber\\
&\quad\qquad
-A^*({\bm k})[1+f_k^2(1-\xi_k^2)]
-A(-{\bm k})[-(1+\xi_k)+(1-\xi_k)^2(\xi_k+3)f_k^2]f_k
\Bigr\}\nonumber\\
&=-{\cal N}_k[A^*({\bm k})a_k+A({\bm k})e^{-i{\bm k}\cdot{\bm d}}b_k]
[1-e^{i{\bm k}\cdot{\bm d}}(1-\xi_k)f_k],
\label{eq:kdduSpsi0}
\end{align}
\end{subequations}
where $A(\bm{k})$ is given in (\ref{eq:Ak}),
\begin{align}
{\cal N}_k&={1\over(1-\alpha)(1-\delta)[1-(\alpha+\beta)]}{\hbar^2\over2m}{f_k\over G_k(d)}\nonumber\\
&={1\over[1-2f_k+(1-\xi_k)(3+\xi_k)f_k^2][1+2f_k+(1-\xi_k)(3+\xi_k)f_k^2][1-f_k^2(1-\xi_k)^2]}
{\hbar^2\over2m}{f_k\over G_k(d)},
\end{align}
and 
\begin{equation}
a_k=1+f_k^2(1-\xi_k^2),\qquad
b_k=[-(1+\xi_k)+(1-\xi_k)^2(\xi_k+3)f_k^2]f_k.
\end{equation}
Here we have made use of the relation
\begin{equation}
A(-{\bm k})=e^{-i{\bm k}\cdot{\bm d}}A({\bm k}).
\end{equation}
It is interesting to observe that the amplitudes in (\ref{eq:kdud-dduSpsi0}) are both given in a factorized form and they are related by the replacement ${\bm d}\leftrightarrow-{\bm d}$.

We are now ready to evaluate the concurrence $C$ in full order.
The relevant matrix elements in (\ref{eq:rhoAB}) are calculated, for a (spatially) long (i.e., an almost monochromatic) incident wave packet $wk_0\gg1$, impinging on a ``small" target $w\gg d$, to be
\begin{subequations}
\begin{align}
a_{11}&=\int_{\Delta\Omega}d^3{\bm k}\,|\langle{\bm k}\downarrow\uparrow\downarrow|S|\psi_0\rangle|^2\nonumber\\
&\sim
{m^2\over2\pi^3\hbar^4w^2}e^{-[{\bm d}^2-({\hat{\bm k}}_0\cdot{\bm d})^2]/8w^2}
|{\cal N}_{k_0}|^2|a_{k_0}-e^{i{\bm k}_0\cdot{\bm d}}b_{k_0}|^2
\int_{\Delta\Omega}d^2\hat{\bm k}\,|1-e^{-ik_0\hat{\bm k}\cdot{\bm d}}(1-\xi_{k_0})f_{k_0}|^2,\\
a_{22}&=\int_{\Delta\Omega}d^3{\bm k}\,|\langle{\bm k}\downarrow\downarrow\uparrow|S|\psi_0\rangle|^2\nonumber\\
&\sim
{m^2\over2\pi^3\hbar^4w^2}e^{-[{\bm d}^2-({\hat{\bm k}}_0\cdot{\bm d})^2]/8w^2}
|{\cal N}_{k_0}|^2|a_{k_0}-e^{-i{\bm k}_0\cdot{\bm d}}b_{k_0}|^2
\int_{\Delta\Omega}d^2\hat{\bm k}\,|1-e^{ik_0\hat{\bm k}\cdot{\bm d}}(1-\xi_{k_0})f_{k_0}|^2,\\
a_{12}&=\int_{\Delta\Omega}d^3{\bm k}\,\langle{\bm k}\downarrow\uparrow\downarrow|S|\psi_0\rangle\langle{\bm k}\downarrow\downarrow\uparrow|S|\psi_0\rangle^*\nonumber\\
&\sim
{m^2\over2\pi^3\hbar^4w^2}e^{-[{\bm d}^2-({\hat{\bm k}}_0\cdot{\bm d})^2]/8w^2}
|{\cal N}_{k_0}|^2(a_{k_0}-e^{i{\bm k}_0\cdot{\bm d}}b_{k_0})
(e^{-i{\bm k}_0\cdot{\bm d}}a_{k_0}^*-b_{k_0}^*)\nonumber\\
&\qquad\qquad\times
\int_{\Delta\Omega}d^2\hat{\bm k}\,e^{ik_0\hat{\bm k}\cdot{\bm d}}
[1-e^{-ik_0\hat{\bm k}\cdot{\bm d}}(1-\xi_{k_0})f_{k_0}]
[1-e^{ik_0\hat{\bm k}\cdot{\bm d}}(1-\xi_{k_0})f_{k_0}]^*.
\end{align}
\end{subequations}
We arrive at the following expressions for the concurrence
\begin{align}
C(\rho)
&={2|a_{12}|\over a_{11}+a_{22}}\nonumber\displaybreak[0]\\
&\sim
{\displaystyle2\bigl|(a_{k_0}-e^{i{\bm k}_0\cdot{\bm d}}b_{k_0})
(a_{k_0}-e^{-i{\bm k}_0\cdot{\bm d}}b_{k_0})\bigr|
\left|\int_{\Delta\Omega}d^2\hat{\bm k}\,e^{ik_0\hat{\bm k}\cdot{\bm d}}
[1-e^{-ik_0\hat{\bm k}\cdot{\bm d}}(1-\xi_{k_0})f_{k_0}]
[1-e^{ik_0\hat{\bm k}\cdot{\bm d}}(1-\xi_{k_0})f_{k_0}]^*
\right|
\over\displaystyle
\bigl|a_{k_0}-e^{i{\bm k}_0\cdot{\bm d}}b_{k_0}\bigr|^2
\int_{\Delta\Omega}d^2\hat{\bm k}\,\bigl|1-e^{-ik_0\hat{\bm k}\cdot{\bm d}}(1-\xi_{k_0})f_{k_0}\bigr|^2
+
\bigl|a_{k_0}-e^{-i{\bm k}_0\cdot{\bm d}}b_{k_0}\bigr|^2
\int_{\Delta\Omega}d^2\hat{\bm k}\,\bigl|1-e^{ik_0\hat{\bm k}\cdot{\bm d}}(1-\xi_{k_0})f_{k_0}\bigr|^2
}
\label{eq:Cinfull}
\end{align}
and the yield 
\begin{align}
P=a_{11}+a_{22}\sim
{m^2\over2\pi^3\hbar^4w^2}e^{-[{\bm d}^2-({\hat{\bm k}}_0\cdot{\bm d})^2]/8w^2}
\bigl|{\cal N}_{k_0}\bigr|^2
\left(
\bigl|a_{k_0}-e^{i{\bm k}_0\cdot{\bm d}}b_{k_0}\bigr|^2
\int_{\Delta\Omega}d^2\hat{\bm k}\,\bigl|1-e^{-ik_0\hat{\bm k}\cdot{\bm d}}(1-\xi_{k_0})f_{k_0}\bigr|^2\right.\qquad\nonumber\\
\left.
+
\bigl|a_{k_0}-e^{-i{\bm k}_0\cdot{\bm d}}b_{k_0}\bigr|^2
\int_{\Delta\Omega}d^2\hat{\bm k}\,\bigl|1-e^{ik_0\hat{\bm k}\cdot{\bm d}}(1-\xi_{k_0})f_{k_0}\bigr|^2
\right)
\label{eq:Pinfull}
\end{align}
in full order.
Observe that the dependence on the incident angle $\theta_0$ remains through the quantity $e^{\pm i{\bm k}_0\cdot{\bm d}}$ even in this case with $w\gg d$, which is nearly absent at the lowest order [comment (i) in~\ref{sssec:lw}].
On the other hand, the dependences of the concurrence $C$ and of the yield $P$ on the scattering angle $\theta_\text{D}$ appear only through the quantity (see Appendix~\ref{app:integralC})
\begin{equation}
{\cal C}\equiv\frac{1}{\Delta\Omega}
\int_{\Delta\Omega}d^2\hat{\bm k}\,e^{ik_0\hat{\bm k}\cdot{\bm d}}
\sim e^{ik_0d\cos\theta_\text{D}}\left[1-{1\over2}(k_0d)^2\sin^2\theta_\text{D}\sin^2(\Delta\theta/2)-ik_0d\cos\theta_\text{D}\sin^2(\Delta\theta/2)\right],
\label{eq:calC}
\end{equation}  
the absolute value of which is nothing but the concurrence at the lowest order (\ref{eq:CBornlw}) and one can further reduce $C$ and $P$ to obtain
\begin{equation}
C(\rho)={|\cal X|\over\cal Y},\qquad
P={m^2\over2\pi^3\hbar^4w^2}\,\Delta\Omega\,e^{-[{\bm d}^2-({\hat{\bm k}}_0\cdot{\bm d})^2]/8w^2}
|{\cal N}_{k_0}|^2{\cal Y},
\label{eqn:CPwithXY}
\end{equation}
where
\begin{subequations}
\label{eqn:XY}
\begin{align}
{\cal X}&\sim2(a_{k_0}-e^{i{\bm k}_0\cdot{\bm d}}b_{k_0})
(a_{k_0}^*-e^{i{\bm k}_0\cdot{\bm d}}b_{k_0}^*)
[
{\cal C}-(1-\xi_{k_0})f_{k_0}-(1-\xi_{k_0})^*f_{k_0}^*+{\cal C}^*|1-\xi_{k_0}|^2|f_{k_0}|^2
],\\
{\cal Y}&\sim
|a_{k_0}-e^{i{\bm k}_0\cdot{\bm d}}b_{k_0}|^2
[
1+|1-\xi_{k_0}|^2|f_{k_0}|^2-{\cal C}(1-\xi_{k_0})^*f_{k_0}^*-{\cal C}^*(1-\xi_{k_0})f_{k_0}
]
\nonumber\\
&\quad
+|a_{k_0}-e^{-i{\bm k}_0\cdot{\bm d}}b_{k_0}|^2
[
1+|1-\xi_{k_0}|^2|f_{k_0}|^2
-{\cal C}(1-\xi_{k_0})f_{k_0}-{\cal C}^*(1-\xi_{k_0})^*f_{k_0}^*
].
\end{align}
\end{subequations}

In Fig.~\ref{fig:CPg}, the concurrence $C$ and the probability $P$ in full order are presented.
Due to the resonant scattering of X between A and B, the concurrence $C$ oscillates as a function of the scattering angle $\theta_\text{D}$ [Fig.~\ref{fig:CPg}(a)].
Such an oscillation is absent at the lowest order in $g_r$ (Fig.~\ref{fig:lwk0hat}).
The resonant scattering can enhance the entanglement, if X is captured in appropriate directions \cite{note:interference}.
The overall characteristics (i)--(iv) of the concurrence $C$ discussed in Sec.~\ref{sec:LowestOrder} for the lowest-order estimation (Fig.~\ref{fig:lwk0hat}) are however more or less kept, apart from the oscillation around the lowest-order values.
As shown in Fig.~\ref{fig:CPDg}(a), the effect of the resonant scattering saturates for $mg_r/\hbar^2d\gg1$.
In this regime, the concurrence $C$ and the probability $P$ in (\ref{eqn:CPwithXY}) are given in terms of 
\begin{subequations}
\label{eqn:XYwithLargeG}
\begin{align}
{\cal X}&\sim2
\left|
1+\frac{e^{2ik_0d}}{k_0^2d^2}
\right|^2\left(
1-
\frac{e^{ik_0d}}{ik_0d}
e^{i{\bm k}_0\cdot{\bm d}}
\right)
\left(
1+
\frac{e^{-ik_0d}}{ik_0d}
e^{i{\bm k}_0\cdot{\bm d}}
\right)
\left(
{\cal C}
-2\frac{\sin k_0d}{k_0d}
+\frac{{\cal C}^*}{k_0^2d^2}
\right),\\
{\cal Y}&\sim
\left|
1+\frac{e^{2ik_0d}}{k_0^2d^2}
\right|^2\left|
1-\frac{e^{ik_0d}}{ik_0d}
e^{i{\bm k}_0\cdot{\bm d}}
\right|^2
\left[
1+\frac{1}{k_0^2d^2}
-2\Im\!\left(
{\cal C}^*
\frac{e^{ik_0d}}{k_0d}
\right)
\right]
\nonumber\\
&\quad
+
\left|
1+\frac{e^{2ik_0d}}{k_0^2d^2}
\right|^2\left|
1-\frac{e^{ik_0d}}{ik_0d}
e^{-i{\bm k}_0\cdot{\bm d}}
\right|^2
\left[
1+\frac{1}{k_0^2d^2}
-2\Im\!\left(
{\cal C}
\frac{e^{ik_0d}}{k_0d}
\right)
\right],
\end{align}
and
\begin{equation}
{\cal N}_{k_0}
\sim\frac{4\pi^2\hbar^4}{3m^2g_rk_0^2}{1\over(1+e^{2ik_0d}/k_0^2d^2)^3}.
\end{equation}
\end{subequations}

Differently from the lowest-order result, the oscillation of the probability $P$ (Young-like interference) is observed in Fig.~\ref{fig:CPg}(b) [cf.\ Eq.\ (\ref{eq:PBorn}) and Sec.~\ref{sec:CharP} for the lowest-order estimation].
This is because the multiple scattering with spin flips breaks the one-to-one correspondence between the path taken by X and the spin state of A and B after the scattering.
Since the two paths are less distinguishable than at the lowest order, they interfere.
As the coupling constant $g_r$ is increased, the probability $P$ for the detection of X in the down state $|\downarrow\rangle_\text{X}$ grows for  $mg_r/\hbar^2d\lesssim1$, while it is suppressed for $mg_r/\hbar^2d\gg1$ [Fig.~\ref{fig:CPDg}(b)].
See (\ref{eqn:XYwithLargeG}) for the $mg_r/\hbar^2d\gg1$ case.
The probability decays like $\sim g_r^{-2}$.

As already mentioned above, the concurrence $C$ and the probability $P$ in full order depend on the incident angle $\theta_0$ even for a large incident wave packet $w\gg d$.
See Figs.~\ref{fig:C-DI} and~\ref{fig:P-DI}.
The dependences are more prominent for a smaller incident wave number $k_0$ and a larger coupling constant $g_r$.

\begin{figure*}
 \begin{center}
\begin{tabular}{lll}
(a)&&(b)\\[-3.5truemm]
\multicolumn{1}{r}{\large$C(\rho)$}&\qquad\qquad\qquad&\multicolumn{1}{r}{\large$Pw^2/\Delta\Omega\,d^2$}\\[-3.5truemm]
  \includegraphics[height=0.31\textwidth]{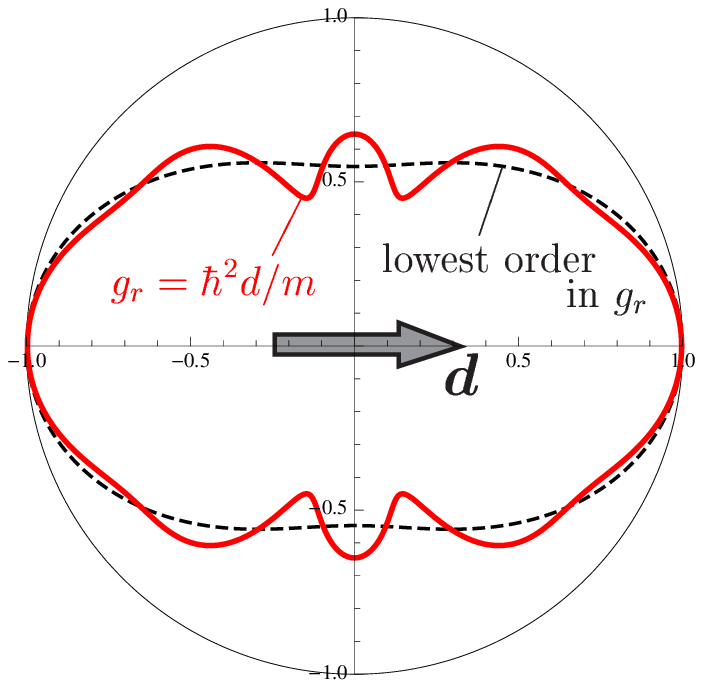}
&&
  \includegraphics[height=0.31\textwidth]{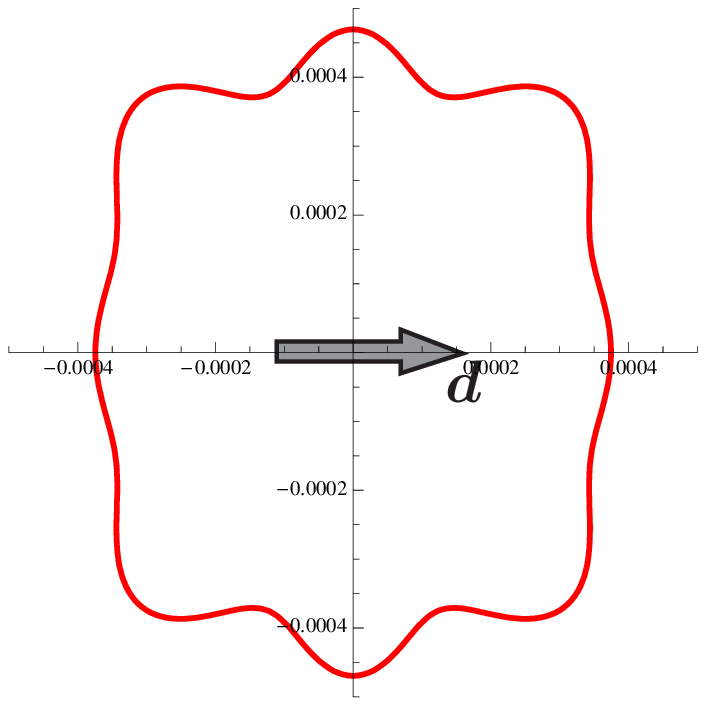}
 \end{tabular}
\end{center}
 \caption{(Color online) (a) Concurrence $C$ and (b) yield $P$ in the full order of perturbation with a large incident wave packet, $w\gg d$, shown in polar coordinates with radii $C$ and $P$ as functions of the scattering angle $\theta_\text{D}$ relative to the alignment $\bm{d}$ of the target qubits A and B\@.
Parameters are: $k_0d=10$, $\theta_0=90^\circ$, $\Delta\theta=\pi/15$, and $mg_r/\hbar^2d=1$.
The corresponding concurrence $C$ in the Born approximation is also shown in (a) as a reference (dashed curve).
}
\label{fig:CPg}
\end{figure*}
\begin{figure*}
 \begin{center}
\begin{tabular}{l@{\qquad\qquad}l}
(a)&(b)\\[-3.5truemm]
  \includegraphics{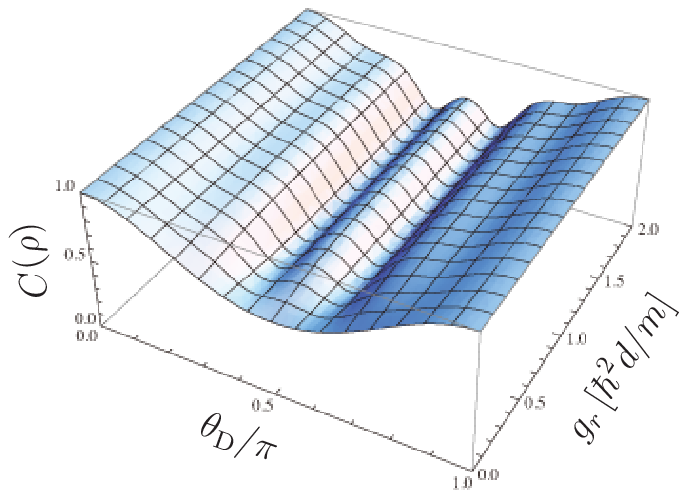}
&
  \includegraphics{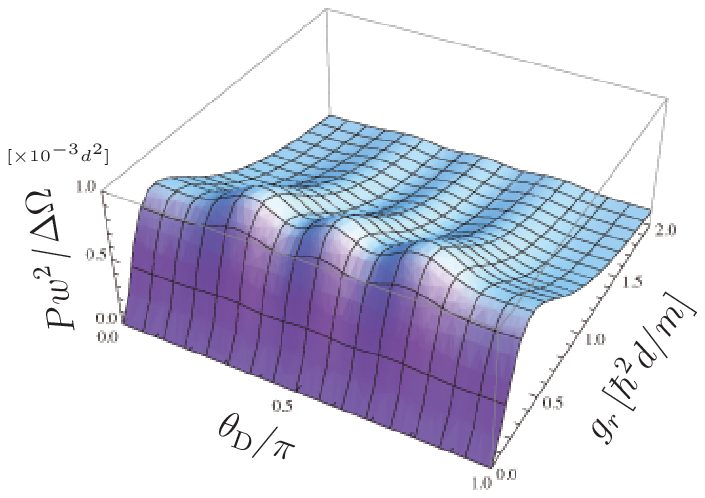}
 \end{tabular}
\end{center}
 \caption{(Color online) (a) Concurrence $C$ and (b) yield $P$ in the full order of perturbation with a large incident wave packet, $w\gg d$, as functions of the scattering angle $\theta_\text{D}$ and the coupling constant $g_r$.
Parameters are: $k_0d=10$, $\theta_0=90^\circ$, and $\Delta\theta=\pi/15$.
}
\label{fig:CPDg}
\end{figure*}
\begin{figure*}
 \begin{center}
\begin{tabular}{l@{\qquad\qquad}l}
  \includegraphics{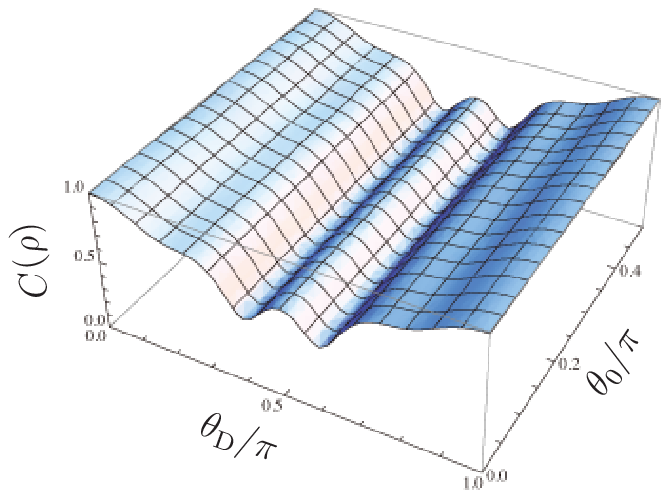}
&
  \includegraphics{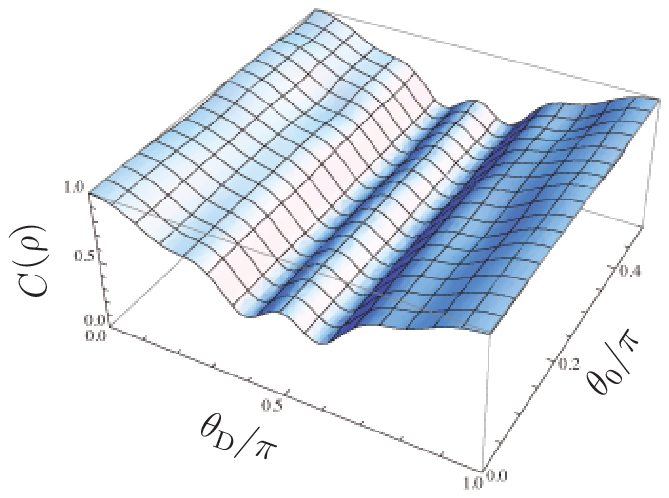}
\\[-50truemm]
(a) $mg_r/\hbar^2d=1$&(b) $mg_r/\hbar^2d=10$\\
\phantom{(a)} $k_0d=10$&\phantom{(b)} $k_0d=10$\\[45truemm]
  \includegraphics{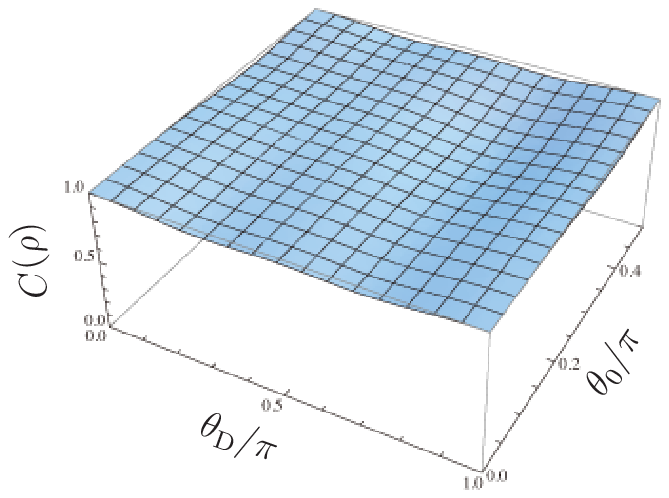}
&
  \includegraphics{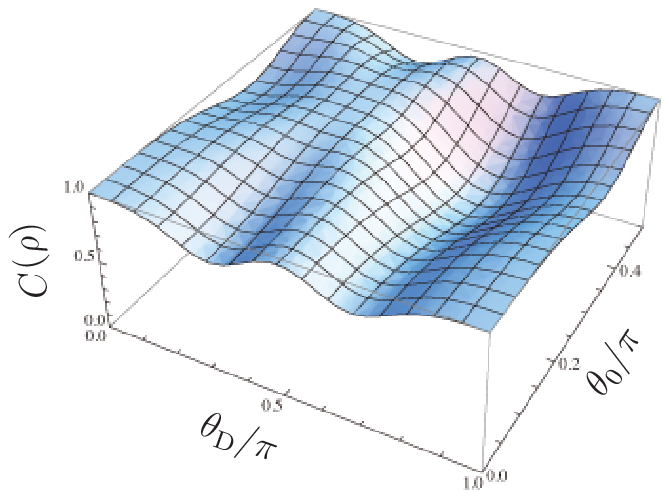}
\\[-50truemm]
(c) $mg_r/\hbar^2d=1$&(d) $mg_r/\hbar^2d=10$\\
\phantom{(c)} $k_0d=\pi$&\phantom{(d)} $k_0d=\pi$\\[43truemm]
 \end{tabular}
\end{center}
 \caption{(Color online) Concurrence $C$ in the full order of perturbation with a large incident wave packet, $w\gg d$, as functions of the scattering angle $\theta_\text{D}$ and the incident angle $\theta_0$.
Parameters are: (a) $k_0d=10$, $mg_r/\hbar^2d=1$; (b) $k_0d=10$, $mg_r/\hbar^2d=10$; (c) $k_0d=\pi$, $mg_r/\hbar^2d=1$; (d) $k_0d=\pi$, $mg_r/\hbar^2d=10$.
$\Delta\theta=\pi/15$ for all cases.
}
\label{fig:C-DI}
\end{figure*}
\begin{figure*}
 \begin{center}
\begin{tabular}{l@{\qquad\qquad}l}
  \includegraphics{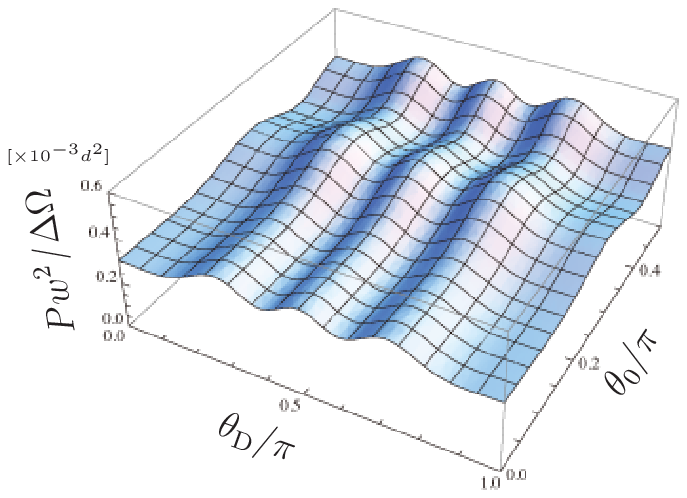}
&
  \includegraphics{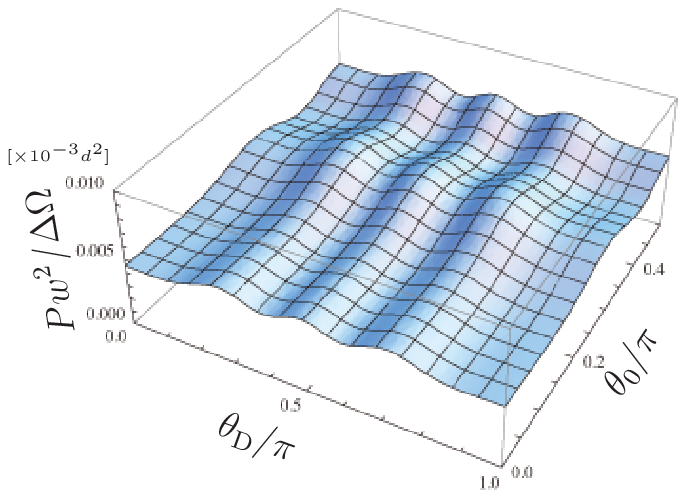}
\\[-50truemm]
(a) $mg_r/\hbar^2d=1$&(b) $mg_r/\hbar^2d=10$\\
\phantom{(a)} $k_0d=10$&\phantom{(b)} $k_0d=10$\\[45truemm]
  \includegraphics{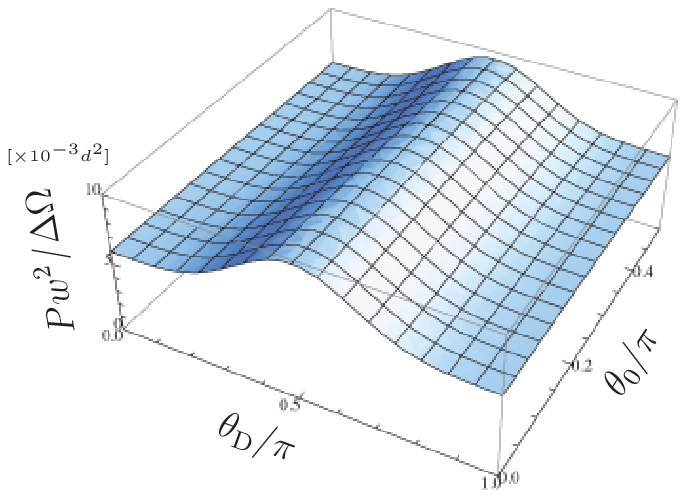}
&
  \includegraphics{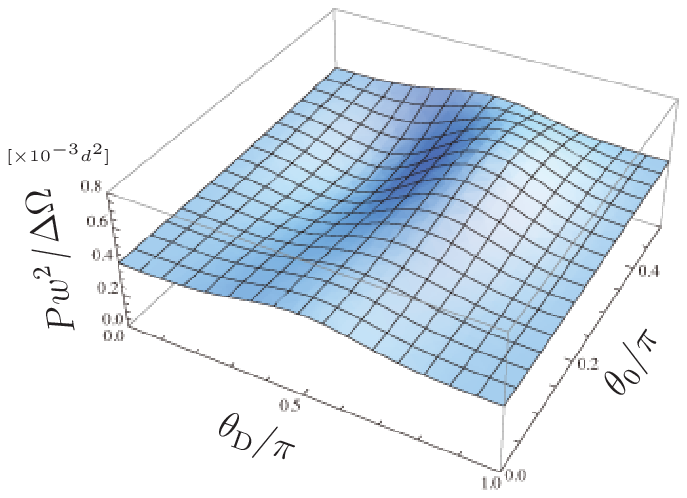}
\\[-50truemm]
(c) $mg_r/\hbar^2d=1$&(d) $mg_r/\hbar^2d=10$\\
\phantom{(c)} $k_0d=\pi$&\phantom{(d)} $k_0d=\pi$\\[43truemm]
\end{tabular}
\end{center}
 \caption{(Color online) Probabilities $P$ corresponding to the concurrences in Fig.~\ref{fig:C-DI}.
}
\label{fig:P-DI}
\end{figure*}

\vfill

\section{Summary and outlook}
\label{sec:summary}
In this paper, a two-qubit system A+B, which is prepared initially in a product state and has no mutual interaction, is shown to be made (maximally) entangled when another mediator qubit X is found spin-flipped in proper directions after scattering.
The interaction between the mediator and each of the target qubits is described by a delta-shaped potential multiplied by the spin-exchange operator.
The scattering (S) matrix elements are calculated both to lowest order and in full order in perturbation to estimate the concurrence $C$ of the target system A+B and its yield $P$.
The concurrence $C$, a measure of the entanglement, is evaluated as a function of such parameters as the momentum $\hbar{\bm k}_0$, width of the wave packet $w$ of the incident qubit, the scattering (detecting) angle $\theta_{\rm D}$ and the detector resolution $\Delta\theta$.
The condition to obtain higher entanglement (concurrence) is derived as a function of these parameters and is interpreted in the context of the (in)distinguishability of alternative paths for quantum particle. 
Even though it is based on the lowest-order expression of the concurrence (\ref{eq:CBorn}) or (\ref{eqn:Cana}), the condition seems to be `globally' valid in full order in perturbation, except `local' oscillations that are due to the presence of multiple scatterings.
In particular, we may conclude that a highest (or maximal) value of concurrence $C$ can be expected in the target system whenever the mediator qubit X is endowed with a short wave length $k_0w\gg 1$ and a spatially long wave packet $w\gg d$ (compared with the size of the target $d$) and is injected and detected by a detector with a higher resolution $\Delta\theta\ll1$ on the line extending the alignment $\bm d$ of the target qubits A and B, that is, $\theta_0=\theta_{\rm D}=0$.

The result, i.e., the maximal entanglement (i.e., the largest concurrence $C=1$) can be obtained in the target qubits A+B whenever the qubit X is injected and captured in the direction $\bm d$ of A+B, is considered to be remarkable since we do not need to adjust any parameters to obtain the maximally entangled state, contrary to the cases with controllable interaction times, where the acquisition of maximal entanglement is conditioned to the proper adjustment of the interaction strength, and to the cases of one-dimensional scattering, where the resonant effects play a crucial role~\cite{ref:qpfep}.
Similar results have also been obtained in the two-dimensional case~\cite{ref:two_dim}.
One could think that in the direction parallel to $\bm d$, the lowest-order contribution, where the path-indistinguishability becomes maximum, resulting in the maximum entanglement, could always overwhelm the higher-order ones reflecting the multiple scatterings.
This does not happen in one-dimensional cases, where there are no other spatial degrees of freedom for the particle to escape, while in spatial dimensions greater than one, the particle could be scattered in other directions than $\bm d$, reducing such resonant effects on the concurrence and keeping the maximal concurrence.

\acknowledgments
This work is partly supported by a Grant-in-Aid for Scientific Research (C) from JSPS, Japan, by a Special Coordination Fund for Promoting Science and Technology from MEXT, Japan, by the the bilateral Italian-Japanese Projects of  MUR, Italy, and by the Joint Italian-Japanese Laboratory  of MAE, Italy. Y. Omar thanks the support from project IT QuantTel and initiatives QuantPrivTel and MMQIRT, and from Funda\c{c}\~{a}o para a Ci\^{e}ncia e a
Tecnologia (Portugal), namely through programs POCTI/POCI/PTDC, partially funded by FEDER (EU).

\appendix
\section{Analytical Estimation of the Concurrence $C$ at the Lowest Order}
\label{app:integralC}
The scattering-angle dependence of the concurrence $C$ and the yield $P$ appears essentially through the quantity
\begin{equation}
{\cal C}\equiv\frac{1}{\Delta\Omega}
\int_{\Delta\Omega}d^2\hat{\bm k}\,e^{ik_0\hat{\bm k}\cdot{\bm d}},
\end{equation}   
the absolute value of which is nothing but the concurrence at the lowest order [see (\ref{eq:CBornlw})] and can be evaluated as follows.
First, the limitation on the polar angle variable $0<\theta<\Delta\theta$, where $\theta$ is measured from the direction of the detector (that is, the scattering direction) $\hat{\bm k}_\text{D}$ and $\Delta\Omega=4\pi\sin^2(\Delta\theta/2)$, is cast into an integration form
\begin{align}
{\cal C}\,\Delta\Omega&=\int_{-\infty}^\infty\frac{d\lambda}{2\pi i}\int d^2\hat{\bm k}\,\frac{e^{i\lambda(\hat{\bm k}\cdot\hat{\bm k}_\text{D}-\cos\Delta\theta)}}{\lambda-i\epsilon}e^{ik_0\hat{\bm k}\cdot{\bm d}}
=\int_{-\infty}^\infty{d\lambda\over2\pi i}{e^{-i\lambda\cos\Delta\theta}\over\lambda-i\epsilon}\int d^2\hat{\bm k}\,e^{i\hat{\bm k}\cdot(k_0{\bm d}+\lambda\hat{\bm k}_\text{D})}.
\intertext{
Now the two-dimensional angle integrations over $\hat{\bm k}$ are trivially done to yield
}
&{-\int_{-\infty}^\infty} d\lambda\,{e^{-i\lambda\cos\Delta\theta}
(e^{i|k_0{\bm d}+\lambda\hat{\bm k}_\text{D}|}-e^{-i|k_0{\bm d}+\lambda\hat{\bm k}_\text{D}|})\over(\lambda-i\epsilon)|k_0{\bm d}+\lambda\hat{\bm k}_\text{D}|}\nonumber\\
&\qquad
=-{1\over k_0d}\int_{-\infty}^\infty d\xi\,{e^{-ik_0d(\xi-\cos\theta_\text{D})\cos\Delta\theta}
(e^{ik_0d\sqrt{\xi^2+\sin^2\theta_\text{D}}}-e^{-ik_0d\sqrt{\xi^2+\sin^2\theta_\text{D}}})\over(\xi-\cos\theta_\text{D}-i\epsilon)\sqrt{\xi^2+\sin^2\theta_\text{D}}},
\label{eq:exct}
\end{align}
where we have changed the integration variable to $\xi=\lambda/(k_0d)+\cos\theta_\text{D}$.
Notice that this expression is exact without any approximation.

Since either exponent of the integrand cannot be well approximated as a quadratic function, in particular for the case of small $\Delta\theta$, in which case the exponent varies quite slowly and higher-order terms could not be neglected, the stationary-phase approximation does not work.
We instead evaluate the integral (\ref{eq:exct}) in the complex $\xi$ plane.
First observe that each term of the integrand has a simple pole at $\xi=\cos\theta_\text{D}+i\epsilon$ in the upper-half plane and two branch points at $\xi=\pm i\sin\theta_\text{D}$.
Draw the two cuts from these branch points, parallel to the real $\xi$ axis, to plus-infinity.
Notice that the quantity $\sqrt{\xi^2+\sin^2\theta_\text{D}}=\sqrt{(\xi+i\sin\theta_\text{D})(\xi-i\sin\theta_\text{D})}$ behaves like 
\begin{equation}
\sqrt{\xi^2+\sin^2\theta_\text{D}}\longrightarrow-|\xi|e^{i\varphi},\quad\varphi={\rm Arg}(\xi+i\sin\theta_\text{D}),
\end{equation}
at infinity $|\xi|\to\infty$, since the argument $\varphi$ is measured anti-clockwise from the positive real axis, $0<\varphi<2\pi$, and then that of $\xi-i\sin\theta_\text{D}$ is expressed as $-(2\pi-\varphi)$ at infinity, except at infinity on the real positive $\xi$ axis, where the arguments are both zero. 
The choice of the arguments is consistent with the fact that the above quantity becomes positive on the entire real $\xi$ axis, where the arguments of $\xi+i\sin\theta_\text{D}$ and $\xi-i\sin\theta_\text{D}$ are $\varphi$ and $-\varphi$ and cancel each other, as they should be.
The argument in the region surrounded by the two cuts in the right-half plane will be scrutinized separately later.
Therefore the first term in (\ref{eq:exct}) can be evaluated by deforming the integration contour downward, for it vanishes at infinity in the lower-half $\xi$ plane.
It is given by the integration along a contour that starts from $+\infty-i\sin\theta_\text{D}$ and runs first to the branch point at $-i\sin\theta_\text{D}$ below the (lower) cut, then to the origin and finally to $+\infty$ on the real positive $\xi$ axis.
It becomes, apart from the factor $-1/k_0d$,
\begin{align}
&\int_0^\infty dx\,{e^{-ik_0d(x-e^{i\theta_\text{D}})\cos\Delta\theta}\over x-e^{i\theta_\text{D}}-i\epsilon}{e^{-ik_0d\sqrt{x(x-2i\sin\theta_\text{D})}}\over\sqrt{x(x-2i\sin\theta_\text{D})}}
+\int_0^\infty dx\,{e^{-ik_0d(x-\cos\theta_\text{D})\cos\Delta\theta}\over x-\cos\theta_\text{D}-i\epsilon}{e^{ik_0d\sqrt{x^2+\sin^2\theta_\text{D}}}\over\sqrt{x^2+\sin^2\theta_\text{D}}}\nonumber\\
&\quad
+i\int_{-\sin\theta_\text{D}}^0dy\,{e^{-ik_0d(iy-\cos\theta_\text{D})\cos\Delta\theta}\over iy-\cos\theta_\text{D}-i\epsilon}{e^{ik_0d\sqrt{\sin^2\theta_\text{D}-y^2}}\over\sqrt{\sin^2\theta_\text{D}-y^2}}.
\end{align}
The second term can be evaluated by similarly deforming the contour but this time upward on the complex $\xi$ plane.
The expression reads, again apart form the same factor $-1/k_0d$, as
\begin{align}
&{-\int_0^\infty dx}\,{e^{-ik_0d(x-e^{-i\theta_\text{D}})\cos\Delta\theta}\over x-e^{-i\theta_\text{D}}-i\epsilon}{e^{ik_0d\sqrt{x(x+2i\sin\theta_\text{D})}}\over\sqrt{x(x+2i\sin\theta_\text{D})}}
-\int_0^\infty dx\,{e^{-ik_0d(x-\cos\theta_\text{D})\cos\Delta\theta}\over x-\cos\theta_\text{D}-i\epsilon}{e^{-ik_0d\sqrt{x^2+\sin^2\theta_\text{D}}}\over\sqrt{x^2+\sin^2\theta_\text{D}}}\nonumber\\
&\quad
+i\int_0^{\sin\theta_\text{D}}dy\,{e^{-ik_0d(iy-\cos\theta_\text{D})\cos\Delta\theta}\over iy-\cos\theta_\text{D}-i\epsilon}{e^{-ik_0d\sqrt{\sin^2\theta_\text{D}-y^2}}\over\sqrt{\sin^2\theta_\text{D}-y^2}}
-2\pi ie^{-ik_0d}\theta(-\cos\theta_\text{D}).
\end{align}
Notice that as for the first terms in the above expressions, representing the contributions coming from integrations along the lower and upper cuts, the integration contours can further be ``pushed'' up (down) by $\sin\theta_\text{D}$ (in other Riemannian sheets), since the argument of $\sqrt{x(x-2i\sin\theta_\text{D})}$ [or $\sqrt{x(x+2i\sin\theta_\text{D})}$] remains negative (or positive) there.
\begin{figure*}
 \begin{center}
  \includegraphics[width=0.48\textwidth]{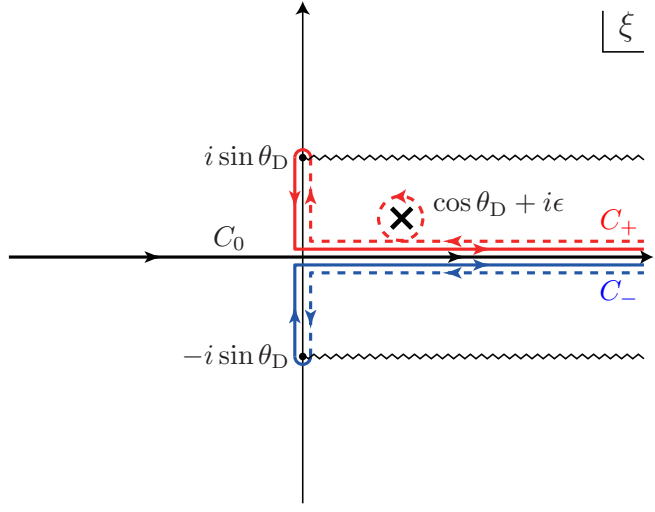}
\end{center}
 \caption{(Color online) Integration contours.  
The original contour $C_0$ and the deformed ones $C_\pm$.}
\label{fig:IntegralContour}
\end{figure*}
That is, we have 
\begin{align}
&\int_0^\infty dx\,{e^{-ik_0d(x-e^{i\theta_\text{D}})\cos\Delta\theta}\over x-e^{i\theta_\text{D}}-i\epsilon}{e^{-ik_0d\sqrt{x(x-2i\sin\theta_\text{D})}}\over\sqrt{x(x-2i\sin\theta_\text{D})}}\nonumber\\
&\qquad
=\int_0^\infty dx\,{e^{-ik_0d(x-\cos\theta_\text{D})\cos\Delta\theta}\over x-\cos\theta_\text{D}-i\epsilon}{e^{-ik_0d\sqrt{x^2+\sin^2\theta_\text{D}}}\over\sqrt{x^2+\sin^2\theta_\text{D}}}
+i\int_{-\sin\theta_\text{D}}^0dy\,{e^{-ik_0d(iy-\cos\theta_\text{D})\cos\Delta\theta}\over iy-\cos\theta_\text{D}-i\epsilon}{e^{-ik_0d\sqrt{\sin^2\theta_\text{D}-y^2}}\over\sqrt{\sin^2\theta_\text{D}-y^2}}
\intertext{(evaluated along the contour $C_-$ in Fig.~\ref{fig:IntegralContour}) and (along the another contour $C_+$)}
&{-\int_0^\infty dx}\,{e^{-ik_0d(x-e^{-i\theta_\text{D}})\cos\Delta\theta}\over x-e^{-i\theta_\text{D}}-i\epsilon}{e^{ik_0d\sqrt{x(x+2i\sin\theta_\text{D})}}\over\sqrt{x(x+2i\sin\theta_\text{D})}}\nonumber\\
&\qquad
=-\int_0^\infty dx{e^{-ik_0d(x-\cos\theta_\text{D})\cos\Delta\theta}\over x-\cos\theta_\text{D}-i\epsilon}{e^{ik_0d\sqrt{x^2+\sin^2\theta_\text{D}}}\over\sqrt{x^2+\sin^2\theta_\text{D}}}
+i\int_0^{\sin\theta_\text{D}}dy\,{e^{-ik_0d(iy-\cos\theta_\text{D})\cos\Delta\theta}\over iy-\cos\theta_\text{D}-i\epsilon}{e^{ik_0d\sqrt{\sin^2\theta_\text{D}-y^2}}\over\sqrt{\sin^2\theta_\text{D}-y^2}}\nonumber\\
&\qquad\quad
+2\pi ie^{ik_0d}\theta(\cos\theta_\text{D}).
\end{align}
We thus obtain the following (still exact) expression for ${\cal C}\,\Delta\Omega$,
\begin{align}
{\cal C}\,\Delta\Omega={}&{-2\pi i}{e^{ik_0d}\over k_0d}\theta(\cos\theta_\text{D})+2\pi i{e^{-ik_0d}\over k_0d}\theta(-\cos\theta_\text{D})\nonumber\\
&{}-{i\over k_0d}\int_{-\sin\theta_\text{D}}^{\sin\theta_\text{D}}dy\,{e^{-ik_0d(iy-\cos\theta_\text{D})\cos\Delta\theta}\over iy-\cos\theta_\text{D}-i\epsilon}{e^{ik_0d\sqrt{\sin^2\theta_\text{D}-y^2}}+e^{-ik_0d\sqrt{\sin^2\theta_\text{D}-y^2}}\over\sqrt{\sin^2\theta_\text{D}-y^2}}.
\end{align}

The last integration over $y$, which is rewritten as an integral over angle $\phi$ through $y=\sin\theta_\text{D}\sin\phi$,
\begin{align}
&{-\frac{i}{k_0d}}\int_{-\pi/2}^{3\pi/2}d\phi\,{e^{ik_0d\sin\theta_\text{D}(\cos\phi-i\cos\Delta\theta\sin\phi)}\over i\sin\theta_\text{D}\sin\phi-\cos\theta_\text{D}}e^{ik_0d\cos\theta_\text{D}\cos\Delta\theta},
\intertext{%
can be cast into an integration along a unit circle $\zeta\equiv e^{i\phi}$ enclosing the origin anti-clockwise on the complex plane, or equivalently that over $z=1/\zeta$ running clockwise,
}
&-{2\over k_0d\sin\theta_\text{D}}\oint dz\,{e^{ik_0d\sin\theta_\text{D}\cos^2(\Delta\theta/2)z}e^{ik_0d\sin\theta_\text{D}\sin^2(\Delta\theta/2)/z}\over[z+\cot(\theta_\text{D}/2)][z-\tan(\theta_\text{D}/2)]}e^{ik_0d\cos\theta_\text{D}\cos\Delta\theta},
\intertext{
and evaluated as the (infinite) sum of the residues within the unit circle, 
}
&2\pi i{e^{ik_0d}\over k_0d}\theta(\cos\theta_\text{D})
-2\pi i{e^{-ik_0d}\over k_0d}\theta(-\cos\theta_\text{D})
\nonumber\\
&\quad
+{4\pi i\over k_0d\sin\theta_\text{D}}
\sum_{n=0}^\infty{[ik_0d\sin\theta_\text{D}\sin^2(\Delta\theta/2)]^{n+1}\over(n+1)!}{1\over n!}\left.
{d^n\over dz^n}
{e^{ik_0d\sin\theta_\text{D}\cos^2(\Delta\theta/2)z}\over[z+\cot(\theta_\text{D}/2)][z-\tan(\theta_\text{D}/2)]}\right|_{z=0}e^{ik_0d\cos\theta_\text{D}\cos\Delta\theta}\nonumber\displaybreak[0]\\
&=2\pi i{e^{ik_0d}\over k_0d}\theta(\cos\theta_\text{D})
-2\pi i{e^{-ik_0d}\over k_0d}\theta(-\cos\theta_\text{D})\nonumber\\
&\quad
-{2\pi i\over k_0d}\sum_{n=0}^\infty\sum_{\ell=0}^n{1\over(n+1)!\ell!}\Bigl\{
[2ik_0d\cos^2(\theta_\text{D}/2)\sin^2(\Delta\theta/2)]^{n+1}[2ik_0d\sin^2(\theta_\text{D}/2)\cos^2(\Delta\theta/2)]^\ell\nonumber\\
&\qquad\qquad\qquad\qquad\qquad\qquad
-[{-2i}k_0d\sin^2(\theta_\text{D}/2)\sin^2(\Delta\theta/2)]^{n+1}[-2ik_0d\cos^2(\theta_\text{D}/2)\cos^2(\Delta\theta/2)]^\ell\Bigr\}\,e^{ik_0d\cos\theta_\text{D}\cos\Delta\theta}.
\end{align}
Therefore, we arrive at
\begin{align}
{\cal C}\,\Delta\Omega=
-{2\pi i\over k_0d}\sum_{n=0}^\infty\sum_{\ell=0}^n&{1\over(n+1)!\ell!}\Bigl\{
[2ik_0d\cos^2(\theta_\text{D}/2)\sin^2(\Delta\theta/2)]^{n+1}[2ik_0d\sin^2(\theta_\text{D}/2)\cos^2(\Delta\theta/2)]^\ell\nonumber\\
&-[-2ik_0d\sin^2(\theta_\text{D}/2)\sin^2(\Delta\theta/2)]^{n+1}[-2ik_0d\cos^2(\theta_\text{D}/2)\cos^2(\Delta\theta/2)]^\ell\Bigr\}\,e^{ik_0d\cos\theta_\text{D}\cos\Delta\theta}.
\end{align}
This is an exact expression, though the double summation seems to be difficult to perform.
On the other hand, it is suited for expansion for small $\Delta\theta$.
Indeed, it allows us to obtain an estimation of ${\cal C}\,\Delta\Omega$, up to, say order $(\Delta\theta)^4$,
\begin{align}
{\cal C}\,\Delta\Omega&\sim-{2\pi i\over k_0d}\left[
2ik_0d\sin^2(\Delta\theta/2)-(k_0d)^2\sin^4(\Delta\theta/4)(2\cos\theta_\text{D}+ik_0d\sin^2\theta_\text{D})
\right]e^{ik_0d\cos\theta_\text{D}\cos\Delta\theta}\nonumber\\
&\sim\Delta\Omega\,e^{ik_0{\bm d}\cdot{\hat{\bm k}}_\text{D}}\left[
1-{1\over2}(k_0d)^2\sin^2\theta_\text{D}\sin^2(\Delta\theta/2)
-ik_0d\cos\theta_\text{D}\sin^2(\Delta\theta/2)
\right],
\end{align}
which is (\ref{eq:calC}) and also yields an approximate expression for the concurrence $C$ to lowest order (\ref{eqn:Cana}),
\begin{equation}
C=|{\cal C}|\sim1-{1\over2}(k_0d)^2\sin^2\theta_\text{D}\sin^2(\Delta\theta/2).
\end{equation}

\section{Projection operators for three-qubit system}
\label{app:PQR}
A quantum system composed of three spin-1/2 particles, say X, A, and B, is decomposed into a spin-3/2 and two spin-1/2 sectors, $1/2\otimes1/2\otimes1/2=3/2\oplus1/2\oplus1/2$.
The spin-3/2 states are totally symmetric with respect to the exchange among three particles, while the spin-1/2 states have mixed symmetries and one of them is anti-symmetric (singlet state) and the other symmetric (triplet state) under the exchange between, say A and B\@.
The projection operators ${\cal P}_s(\gamma)$, ${\cal Q}_s(\gamma)$, and ${\cal R}_\pm$ introduced in Sec.~\ref{ssedc:multiplescattering} are related with these spin decompositions.

First of all, since the square of the total spin is expressed as
\begin{equation}
{\bm S}_\text{XAB}^2={\hbar^2\over4}\Bigl({\bm\sigma}^\text{(X)}+{\bm\sigma}^\text{(A)}+{\bm\sigma}^\text{(B)}\Bigr)^2
={\hbar^2\over4}\Bigl(9+2{\bm\sigma}^\text{(X)}\cdot({\bm\sigma}^\text{(A)}+{\bm\sigma}^\text{(B)})+2{\bm\sigma}^\text{(A)}\cdot{\bm\sigma}^\text{(B)}\Bigr),
\end{equation}
the operator ${\cal R}_+$ extracts the total spin 3/2 states, while ${\cal R}_-$ is a projection upon the four-dimensional subspace with the total spin 1/2.
That is,
\begin{equation}
{\cal R}_+=1\;\Leftrightarrow\;\hbox{\rm total spin }{3\over2},\qquad
{\cal R}_+=0\;\Leftrightarrow\;\hbox{\rm total spin }{1\over2}
\end{equation}
or
\begin{equation}
{\cal R}_-=1\;\Leftrightarrow\;\hbox{\rm total spin }{1\over2},\qquad
{\cal R}_-=0\;\Leftrightarrow\;\hbox{\rm total spin }{3\over2}.
\end{equation}

Since ${\cal Q}_s(\gamma)$ is orthogonal to ${\cal R}_+$, while the subspace projected by the latter is included by that by ${\cal P}_s(\gamma)$ [see Eq.\ (\ref{eq:P3})], the projection operator ${\cal P}_s(\gamma)$ extracts all spin-3/2 states and a part of spin-1/2 states.
Indeed, it is not difficult to show that the latter sector with the total spin 1/2 (belonging to ${\cal R}_-=1$ sector) are spanned by the following eigenvectors
\begin{equation}
|\psi_+\rangle=\sqrt{1-p}|\phi_+\rangle_\text{XAB}-is\sqrt{p}|\uparrow\rangle_\text{X}|0,0\rangle_\text{AB},
\quad
|\psi_-\rangle=\sqrt{1-p}|\phi_-\rangle_\text{XAB}-is\sqrt{p}|\downarrow\rangle_\text{X}|0,0\rangle_\text{AB},
\quad
p={3\over4}-3\gamma,
\end{equation}
where states $|\phi_\pm\rangle_\text{XAB}$ are particular states belonging to the AB-triplet subspace
\begin{align}
|\phi_+\rangle_\text{XAB}
&=\sqrt{1\over3}|\uparrow\rangle_\text{X}|1,0\rangle_\text{AB}-\sqrt{2\over3}|\downarrow\rangle_\text{X}|\uparrow\uparrow\rangle_\text{AB}={1\over\sqrt3}\Bigl(|\uparrow\rangle_\text{A}|0,0\rangle_\text{XB}+|\uparrow\rangle_\text{B}|0,0\rangle_\text{XA}\Bigr),\\
|\phi_-\rangle_\text{XAB}
&=\sqrt{2\over3}|\uparrow\rangle_\text{X}|\downarrow\downarrow\rangle_\text{AB}-\sqrt{1\over3}|\downarrow\rangle_\text{X}|1,0\rangle_\text{AB}={1\over\sqrt3}\Bigl(|\downarrow\rangle_\text{A}|0,0\rangle_\text{XB}+|\downarrow\rangle_\text{B}|0,0\rangle_\text{XA}\Bigr),
\end{align}
and $|0,0\rangle_\text{XA}$ the singlet XA-state, and so on.
That is, wighted superpositions of the AB-triplet and AB-singlet states, characterized by the parameter $p$ (or $\gamma$), as well as the total spin 3/2 states, are extracted by the projection operator ${\cal P}_s(\gamma)$.
\end{widetext}


\begin{thebibliography}{10}
\bibitem{note:interference}
The concurrence can be degraded by the multiple scatterings, if X is captured in inappropriate directions.
Oscillation is also observed in the one-dimensional case.
See the discussion in Ref.\ \cite{ref:qpfep}.
\end{thebibliography}
\end{document}